\documentclass[12pt]{article}
\pdfoutput=1
\usepackage[height=8.85in,width=6.45in]{geometry}

\usepackage[utf8]{inputenc}
\usepackage{amsmath,amssymb,amsthm,mathtools}
\usepackage{times,courier,bm,booktabs,array,graphicx}
\usepackage[svgnames,dvipsnames]{xcolor}
\usepackage[colorlinks,citecolor=DarkGreen,linkcolor=FireBrick,urlcolor=FireBrick,linktocpage,unicode,psdextra]{hyperref}
\usepackage{cite}
\usepackage{enumitem}
\usepackage{slashed}
\usepackage{dsfont}
\usepackage[bottom]{footmisc}
\usepackage{tikz}
\usetikzlibrary{arrows.meta,positioning,calc,decorations.markings}

\newcommand{\bea}{\begin{eqnarray}}
\newcommand{\eea}{\end{eqnarray}}
\numberwithin{equation}{section}
\setlist[itemize]{leftmargin=1.5em,itemsep=2pt,topsep=4pt}
\setlist[enumerate]{leftmargin=1.7em,itemsep=2pt,topsep=4pt}

\newcommand{\R}{\mathbb{R}}
\newcommand{\Z}{\mathbb{Z}}

\newcommand{\II}{\mathrm{II}}
\newcommand{\osp}{\mathfrak{osp}}
\newcommand{\so}{\mathfrak{so}}
\newcommand{\g}{\mathfrak{g}}
\newcommand{\pd}{\partial}
\newcommand{\bz}{\bar z}
\newcommand{\Res}{\mathop{\mathrm{Res}}}

\begin{document}
\begin{titlepage}
\begin{flushright}
KYUSHU-HET-386
\end{flushright}
\vskip 2.5cm
\begin{center}
{\Large\bfseries The $(5,5)$ String}
\vskip 1.2cm
Justin Kaidi
\vskip .8cm
 Department of Physics, Kyushu University, Fukuoka 819-0395, Japan\\
 Institute for Advanced Study, Kyushu University, Fukuoka 819-0395, Japan\\
 Quantum and Spacetime Research Institute (QuaSR), Kyushu University, Fukuoka 819-0395, Japan\\
 \vskip 0.1 in
\texttt{kaidi.justin@phys.kyushu-u.ac.jp}

\end{center}
\vskip 1.2cm
\noindent
\begin{quote}
We study an $N=(2,1)$ heterotic string with right-movers on $\R^{2,2}$ and left-movers of signature $(5,5)$, obtained by adjoining the lattice $\II_{3,3}$. Unlike the standard $N=(2,1)$ strings it requires no null current, and its target space remains four-dimensional. All of its states, an infinite tower with sixteen supercharges, are massless. For the neutral bosons, the Kutasov-Martinec determinant action is dual to a cubic action for a self-dual metric coupled to self-dual Maxwell fields, and we complete it with fermions using a single superfield. For the charged states we give evidence that the couplings are the structure constants of a Lie superalgebra formed by the physical states of the left-movers. We are thus led to propose that the classical target theory is a self-dual gauge theory for this superalgebra: every state of the string is a gauge field, and gravity is part of the gauge algebra. This would be a purely cubic field theory of all the modes of a closed string.
\end{quote}
\end{titlepage}
\setcounter{tocdepth}{2}

\tableofcontents

\section{Introduction}
\label{sec:intro}

Among the known string theories, those with $N=2$ supersymmetry on the worldsheet are in many ways the simplest. The critical dimension is four, the signature of spacetime is $(2,2)$, and the spectrum consists of a single massless state. As shown by Ooguri and Vafa \cite{Ooguri:1990ww,Ooguri:1991fp}, the target-space theory of the closed $N=2$ string is self-dual gravity: the one field is a deformation of the K\"ahler potential, its equation of motion is the Pleba\'nski equation \cite{Plebanski:1975wn}, the three-point amplitude is nonzero, and all higher tree-level amplitudes vanish. The theory is thus a string theory which is at the same time an integrable field theory, and it has served as a laboratory for many ideas, from topological strings \cite{Berkovits:1994vy} to self-dual Yang-Mills theory and its supersymmetric extensions \cite{Siegel:1992xp,Siegel:1992wd,Chalmers:2000bh}.

A richer set of theories is obtained by taking only the right-movers to have $N=2$ supersymmetry, while the left-movers are those of the bosonic string or of the $N=1$ string. These $N=(2,0)$ and $N=(2,1)$ heterotic strings were introduced by Ooguri and Vafa \cite{Ooguri:1991ie}, who found that the left-movers must live in a spacetime with \emph{two} times, of signature $(26,2)$ or $(10,2)$, and that consistency then requires gauging an additional current $J=v\cdot\pd X$ built from a null vector $v$. Some years later, Kutasov and Martinec made the striking proposal that the $N=(2,1)$ string is a common origin of the worldvolume theories of strings and membranes \cite{Kutasov:1996fp,Kutasov:1996zm,Kutasov:1996vh}. Depending on the direction of $v$ the target space is $1+1$ or $2+1$ dimensional, the target-space fields are the embedding coordinates of a brane in ten or eleven dimensions together with their superpartners, and the exact classical action is of Nambu-Goto/Born-Infeld type. Different string theories, as well as the eleven-dimensional membrane, then arise at different points in the moduli space of a single $N=2$ heterotic string. This subject was largely dormant for many years,\footnote{See however \cite{Rabinovici:1997hd,Hull:1997kk,Lawrence:1996ti,Junemann:1997rx} and \cite{Martinec:1996wn,deBoer:1997lpg,Rudychev:1997ui} for a sample of the early work which followed.} but has recently been revisited \cite{Giveon:2024fhz}, and the self-dual theories to which it is related have in the meantime come to play a central role in twistor and celestial approaches to scattering amplitudes \cite{Costello:2021bah,Costello:2022wso,Costello:2022upu,Bittleston:2022nfr,Bittleston:2022jeq}.

In all of this work a central role is played by the null vector $v$. Its direction relative to the base $\R^{2,2}$ and to the internal directions decides what the target space is, and yet it is not determined by anything---it is put in by hand, breaks the symmetries of the two-time spacetime in which the left-movers live, and has no dynamics of its own. As emphasized already in \cite{Kutasov:1996vh}, the geometrical role of the null reduction is a persistent puzzle. The goal of this paper is to shed some light on this puzzle. We will do so in two steps. The first is to identify the algebraic structure of which the null vector is a part. The second, which is a consequence of the first, is to observe that there is a member of the family of $N=(2,1)$ strings which needs no null vector at all, and to study this string in some detail. This study turns out to have a payoff which goes beyond the original puzzle. It is an old idea that the states of a string form a Lie algebra, and that this algebra acts as a symmetry of the theory \cite{Borcherds:1983sq,Harvey:1995fq}; self-dual gauge theory based on such an algebra was proposed in \cite{Kutasov:1996vh}. We will find that in the string without a null vector this idea is realized rather explicitly, with the graviton among the generators. Since much of the relevant background is no longer widely known, we will try to be self-contained throughout.

\subsection{The basic idea }
 Gauging the null current $J=v\cdot\pd X$ and its superpartner $\Psi=v\cdot\psi$ introduces a pair of anticommuting ghosts $(b',c')$ and a pair of commuting ghosts $(\beta',\gamma')$. Our starting point is the observation that these ghosts have exactly the free-field content of two odd directions of spacetime. In other words, the twelve bosons $X^M$ and fermions $\psi^M$ of the left-movers, together with the ghosts of the null current, form an $N=1$ string on the superspace $\R^{10,2|2}$, whose symmetry is the orthosymplectic algebra $\osp(10,2|2)$. The BRST charge of the null current,
\bea
\label{eq:introqv}
Q_v=\oint\big(c'\,v\!\cdot\!\pd X+\gamma'\,v\!\cdot\!\psi\big)~,
\eea
is an odd element of this algebra, and it is nilpotent precisely when $v^2=0$. Gauging the null current then amounts to passing to the cohomology of $Q_v$, which removes two even and two odd directions and leaves a string on $v^\perp/v=\R^{9,1}$. This is an instance of what is known in the literature on sigma models with supergroup symmetry as \emph{cohomological reduction} \cite{Candu:2010yg}; we will explain this notion from scratch in Section \ref{sec:mechanism}. From this point of view, a choice of null vector is a choice of an odd nilpotent element of $\osp(10,2|2)$, and the different vacua of the $N=(2,1)$ string are the different orbits of such elements.

The reduction just described makes sense in other signatures. For the twelve even directions there are three signatures relevant for strings,
\bea
\label{eq:introrealforms}
\R^{10,2|2}\to\R^{9,1}~, \hspace{0.4 in} \R^{6,6|2}\to\R^{5,5}~, \hspace{0.4 in} \R^{2,10|2}\to\R^{1,9}~,
\eea
the first of which is the standard $N=(2,1)$ string. They are distinguished by an invariant $\sigma\in8\Z$, the signature of the internal lattice, which is not changed by the reduction. The second real form is special, since it is the only one for which the null vector can be taken to lie entirely in the internal, purely left-moving directions. This gives rise to a string without a null current whose right-movers are those of the $N=2$ string on $\R^{2,2}$, and whose left-movers are those of the ordinary $N=1$ string on
\bea
\label{eq:introtarget}
\R^{2,2}\times\II_{3,3}~,
\eea
where $\II_{3,3}$ is the even self-dual lattice of signature $(3,3)$. The left-movers thus see a spacetime of signature $(5,5)$, and we will refer to this theory as \emph{the $(5,5)$ string}.\footnote{Note that $(5,5)$ here does not refer to the worldsheet supersymmetry, which will be $(2,1)$ unless otherwise stated. The author apologizes in advance for the inevitable confusion this will bring. He is open to suggestions for a better name. } Its target space is the full $\R^{2,2}$, with no reduction to fewer dimensions. Moreover, because the lattice is indefinite, the oscillator level of a state can be compensated by the negative norm of its charge. As a result \emph{every} state of the string is massless on $\R^{2,2}$, and the spectrum is an infinite tower of massless states of arbitrarily high helicity, with sixteen supercharges.

The bulk of this paper is devoted to the fascinating properties of this string. We find the following,
\begin{itemize}

\item \textbf{The bosonic theory and its dual.} The neutral spectrum contains eight bosons and eight fermions. Starting from the Kutasov-Martinec determinant action with six scalars of signature $(3,3)$, the duality of \cite{AbouZeid:1997he} recasts it as a very simple cubic field theory, whose equations describe a self-dual metric, six self-dual Maxwell fields, and a scalar sourced by the Maxwell fields.

\item \textbf{A superfield completion.} The geometric equations admit a completion in a superfield $\Phi$ with four odd coordinates obeying
\bea
\label{eq:introsuperpleb}
\det\big(\pd_a\pd_{\bar b}\Phi\big)=-1~.
\eea
The superfield component action reproduces the dual bosonic theory, and its cubic couplings match those of the neutral string states. We verify the vanishing of its four- and five-point tree amplitudes; the vanishing for any number of legs follows from the general argument of Section \ref{sec:algebra}. These results support a description of the neutral sector by a chiral $N=4$ self-dual gravity theory.

\item \textbf{An algebra organizing the charged tower.} In an ordinary heterotic string, the left-moving currents generate the gauge algebra. In the $(5,5)$ string every physical state of the left-movers plays the role of a current, and we give evidence that the couplings of the charged states are the structure constants of the resulting Lie superalgebra $\g$. For charges which are null and mutually orthogonal we find the bracket explicitly: it is the Poisson bracket of the neutral sector, deformed by the charges. For other charges the bracket of two massless states can be an excited state of the string, so that the higher levels are needed for the algebra to close. We propose that $\g$, together with the kinematics of the right-movers, determines the complete target theory, which is cubic. The proposed target theory is given in  (\ref{eq:TOE}). Given the Jacobi identity of $\g$, all of its tree amplitudes beyond three points vanish.
\end{itemize}

\subsection{Broader lessons}

These results suggest several broader lessons about string theory which go beyond the particular model. 
For easy reference, we list these in bullet form here, returning to them in more detail in Section \ref{sec:discussion},
\begin{itemize}
\item \textbf{The states of the string generate its symmetry.} In the proposed target theory there is a single field, taking values in a single Lie superalgebra $\g$, with a single cubic interaction. The graviton, the Maxwell fields, the scalar, their fermionic partners, the charged multiplets, and every excited state of the string are components of this field, and every coupling among them is a structure constant of $\g$. In other words, the whole theory is a self-dual gauge theory for $\g$, in which each state of the string is the gauge field of its own generator. In an ordinary string theory an algebra of this kind, if present, is obscured by the masses of the excited states. Here every state is massless, and the algebra can be read off directly from the three-point amplitudes. 

\item \textbf{Gravity is part of the gauge algebra.} The neutral part of $\g$ is a Poisson algebra, whose even elements generate the area-preserving diffeomorphisms of a plane, and which produces self-dual gravity when combined with the right-moving kinematics. Gravity is therefore not coupled to a gauge theory as a separate sector, but is rather the neutral part of the gauge algebra itself. Conversely, the charged states cannot be separated from gravity or from one another. The bracket of two states of opposite charge is neutral, and nonorthogonal null charges generate timelike charges, which require excited string states for closure.

\item \textbf{A closed string can have a cubic field theory.} A field which takes values in a space of states of the string, with a cubic interaction defined by an algebraic operation on that space, is what the open string field theory of Witten provides \cite{Witten:1985cc}. For closed strings the situation is ordinarily more involved, with the covariant theory being non-polynomial and having new elementary vertices at every order \cite{Zwiebach:1992ie}. The proposal above is, in effect, a field theory of all the modes of a closed string whose interaction is purely cubic and is given by a Lie bracket. That self-dual strings admit simple field theories of this kind is familiar when there is a single field \cite{Ooguri:1990ww,Ooguri:1991fp}, and for open $N=2$ strings a string field theory using Witten's product was formulated in \cite{Berkovits:1997pq}. Here the same simplicity appears to persist for the infinite tower arising from the left-moving oscillators and the indefinite internal lattice. The comparison has clear limits, though, since our field takes values in the physical states of the left-movers and not in the full off-shell state space of the string.

\item \textbf{The null vector is a choice, and it hides this structure.} From the present point of view the null vector is not a fundamental ingredient of the $N=(2,1)$ string. It is an odd nilpotent element of an orthosymplectic algebra whose choice selects a vacuum, and in one real form it can be dispensed with altogether. This is the member of the $N=(2,1)$ family which makes the above structure as manifest as possible.
\end{itemize}

\paragraph{Organization.} This paper is organized as follows. In Section \ref{sec:review} we review the $N=2$ string and its heterotic versions, assuming no prior familiarity with either. In Section \ref{sec:mechanism} we explain cohomological reduction and show that the gauging of the null current is an example of it. In Section \ref{sec:model} we define the $(5,5)$ string and determine its spectrum and supersymmetry. Section \ref{sec:neutral} is devoted to the neutral sector and its description in terms of a single superfield, and Section \ref{sec:charged} to the charged sector and the Lie superalgebra $\g$. In Section \ref{sec:vacua} we describe the other vacua of the reduction, and in Section \ref{sec:discussion} we discuss the broader picture and possible connections to ghost branes and two-time physics. Our conventions are collected in Appendix \ref{app:conventions}, and the zero modes of the null ghosts are discussed in Appendix \ref{app:zeromodes}. The vertex operators and correlation functions used in the text are given in Appendix \ref{app:correlators}, and the states of the first massive level in Appendix \ref{app:levelone}.

\section{$N=2$ strings and their heterotic versions}
\label{sec:review}

We begin by reviewing the $N=2$ string and the $N=2$ heterotic strings. Nothing in this section is new. Our aim is to collect, in one place and in a uniform notation, the facts that will be needed later, and to explain where the null vector comes from. Readers familiar with \cite{Ooguri:1991fp,Ooguri:1991ie,Kutasov:1996vh} may wish to skip to Section \ref{sec:reviewsummary}, where the basic facts used in the rest of the paper are summarized.

\subsection{The $N=2$ string}
\label{sec:ntwostring}

The bosonic string is obtained by gauging the Virasoro algebra on the worldsheet, and the superstring by gauging its $N=1$ supersymmetric extension. The $N=2$ string is obtained by gauging the $N=2$ superconformal algebra, whose generators are the stress tensor $T$, two supercurrents $G^\pm$ of weight $3/2$, and a $U(1)$ current $J$ of weight one, under which $G^\pm$ have charge $\pm1$. Gauging each generator introduces a pair of ghosts: the usual $(b,c)$ ghosts for $T$, with central charge $-26$; a pair of commuting ghosts for each of $G^\pm$, with central charge $+11$ each; and a pair of anticommuting ghosts of weights $(1,0)$ for $J$, with central charge $-2$. The total ghost central charge is
\bea
\label{eq:ntwoghosts}
c_{\mathrm{gh}}=-26+11+11-2=-6~,
\eea
so the matter theory must have $c=6$. The simplest way to realize the $N=2$ algebra is with a complex boson and a complex fermion, which together have $c=3$. The critical $N=2$ string thus has two complex dimensions, or four real ones.

To write the generators, let $x^\mu$ and $\psi^\mu$, $\mu=1,\dots,4$, be real bosons and fermions, with a flat metric $\eta_{\mu\nu}$. The $N=2$ algebra requires in addition a complex structure, that is an antisymmetric tensor $I_{\mu\nu}$ with $I^\mu{}_\nu I^\nu{}_\lambda=-\delta^\mu_\lambda$. Up to normalization, the generators are then
\bea
\label{eq:ntwogenerators}
T=-{1 \over 2}\,\pd x\!\cdot\!\pd x-{1 \over 2}\,\psi\!\cdot\!\pd\psi~, \hspace{0.3 in} G^\pm=\big(\eta_{\mu\nu}\pm i\,I_{\mu\nu}\big)\psi^\mu\pd x^\nu~, \hspace{0.3 in} J={i \over 2}\,I_{\mu\nu}\psi^\mu\psi^\nu~.
\eea
In complex coordinates $z^a$, $a=1,2$, adapted to $I$, the metric is a hermitian form $\eta_{a\bar b}$ and $G^\pm$ involve only the holomorphic (resp. antiholomorphic) fermions.

Each complex coordinate consists of two real directions of the same sign, so the signature of the metric is $(4,0)$, $(2,2)$, or $(0,4)$. As we will see momentarily, the physical states are massless. In Euclidean signature the condition $k^2=0$ has no nonzero real solutions, so to have propagating states we must take the signature to be $(2,2)$. This is the first unusual feature of the $N=2$ string: \emph{spacetime has two times}. For a fixed $I$, the manifest Lorentz symmetry is its stabilizer $U(1,1)\subset SO(2,2)$.\footnote{As discussed in \cite{Kutasov:1996vh}, one may think of this breaking as spontaneous, with $I_{\mu\nu}$ the expectation value of a target-space field. A formulation with manifest $\mathrm{SO}(2,2)$ covariance is obtained from the $N=4$ string \cite{Siegel:1992ev}.}

Let us now turn to the spectrum. In the bosonic string, the Virasoro constraints remove the oscillators along two of the 26 directions, leaving 24 transverse ones. In the $N=2$ string the constraints are stronger. Indeed, the generators (\ref{eq:ntwogenerators}) remove the oscillators along two \emph{complex} directions, which is all of them. At nonzero momentum, the relative BRST cohomology consequently contains a single massless state in each picture, represented by a scalar $\varphi$ with $k^2=0$ \cite{Junemann:1997rx}. The zero-momentum cohomology has additional picture-dependent structure, which is not included in this propagating spectrum. For the closed string, with $N=2$ supersymmetry for both left- and right-movers, the tree-level amplitudes of this scalar were computed in \cite{Ooguri:1990ww,Ooguri:1991fp}. The three-point amplitude is
\bea
\label{eq:closedthree}
A_3=\big(k_1Ik_2\big)^2~, \hspace{0.4 in} k_1Ik_2\equiv I^{\mu\nu}k_{1\mu}k_{2\nu}~,
\eea
with one factor of $k_1Ik_2$ coming from the left-movers and one from the right-movers, while the four-point amplitude vanishes identically. It is expected that all higher-point tree amplitudes vanish as well. These amplitudes are reproduced by the field theory with action
\bea
\label{eq:plebanskiaction}
S=\int d^4x\,\Big[{1 \over 2}\,\eta^{a\bar b}\pd_a\varphi\,\pd_{\bar b}\varphi+{1 \over 3}\,\varphi\,\epsilon^{ac}\epsilon^{\bar b\bar d}\,\pd_a\pd_{\bar b}\varphi\,\pd_c\pd_{\bar d}\varphi\Big]~,
\eea
whose equation of motion is equivalent, for a suitable normalization of $\varphi$, to the statement that the K\"ahler metric $g_{a\bar b}=\eta_{a\bar b}+\pd_a\pd_{\bar b}\varphi$ has the same determinant as the flat one. This is the Pleba\'nski equation \cite{Plebanski:1975wn}. It says that $g_{a\bar b}$ is Ricci-flat, which for a four-dimensional K\"ahler metric implies self-dual curvature, with a choice of orientation. The closed $N=2$ string is thus a theory of self-dual gravity, with $\varphi$ a deformation of the K\"ahler potential. Likewise, the open $N=2$ string describes self-dual Yang-Mills theory.

Two features of this result will recur throughout this paper. The first is that the theory is cubic, with no amplitudes beyond three points; this is a reflection of the integrability of the self-duality equations. The second is that the three-point amplitude (\ref{eq:closedthree}) is the square of an antisymmetric quantity, $k_1Ik_2$, which is the structure constant of a Lie algebra. Since this algebra will play a central role, let us describe it explicitly. A \emph{Poisson algebra} is a space of functions with a bracket $\{f,g\}$ which is antisymmetric and obeys the Jacobi identity and the Leibniz rule $\{f,gh\}=\{f,g\}h+g\{f,h\}$, the standard example being the Poisson bracket of functions on a phase space. The tensor $I$ defines such a bracket on the functions on $\R^{2,2}$,
\bea
\label{eq:poisson}
\{f,g\}=I^{\mu\nu}\,\pd_\mu f\,\pd_\nu g~, \hspace{0.4 in} \{e^{ik_1\cdot x},e^{ik_2\cdot x}\}=-\big(k_1Ik_2\big)\,e^{i(k_1+k_2)\cdot x}~,
\eea
so that $k_1Ik_2$ is its structure constant in the basis of plane waves. With this bracket the functions form an infinite-dimensional Lie algebra, in which a function $f$ generates the transformation $\delta x^\mu=I^{\mu\nu}\pd_\nu f$ of the coordinates. For functions on a plane, these transformations are the area-preserving diffeomorphisms. In Sections \ref{sec:neutral} and \ref{sec:charged} we will also meet the Poisson brackets of functions of the two holomorphic coordinates alone, and of the two antiholomorphic coordinates alone, whose structure constants are the quantities $X$ and $\bar X$ of (\ref{eq:xdef}).\footnote{The structure of self-dual gravity amplitudes as squares of those of self-dual Yang-Mills, with the Poisson algebra as ``kinematic algebra'', has been much studied in recent years; see e.g. \cite{Monteiro:2011pc}.}

\subsection{$N=2$ heterotic strings and the null current}
\label{sec:heteroticreview}

In the ordinary heterotic string, the right-movers are those of the $N=1$ string and the left-movers those of the bosonic string. Since the two have different critical dimensions, ten and twenty-six, sixteen of the left-moving bosons have no right-moving partners, and consistency requires them to live on an even self-dual lattice. The $N=2$ heterotic strings of Ooguri and Vafa \cite{Ooguri:1991ie} are built in the same way, with the right-movers now those of the $N=2$ string. The left-movers can be those of the bosonic string, giving the $N=(2,0)$ string, or those of the $N=1$ string, giving the $N=(2,1)$ string. It is the latter which will concern us here.

An immediate problem arises. The right-movers require a base of signature $(2,2)$, shared with the left-movers. On the other hand, the usual ghost-free $N=1$ string has critical dimension ten and signature $(9,1)$, and a space with one time has no subspace of signature $(2,2)$. The resolution proposed in \cite{Ooguri:1991ie} is that the left-movers should live in \emph{twelve} dimensions, with signature $(10,2)$,
\bea
\label{eq:tentwo}
\R^{10,2}=\R^{2,2}\times\R^{8}~,
\eea
with the eight extra directions purely left-moving and compactified on an even self-dual lattice, for example that of $E_8$. For this to be a critical string, something must reduce the central charge by the equivalent of two dimensions, and something must remove the negative-norm states associated with the second time. Both are achieved by enlarging the left-moving gauge algebra by a $U(1)$ current and its superpartner,
\bea
\label{eq:nullcurrent}
J=v\!\cdot\!\pd X~, \hspace{0.4 in} \Psi=v\!\cdot\!\psi~,
\eea
where $v$ is a constant vector in $\R^{10,2}$. Gauging $J$ and $\Psi$ introduces anticommuting ghosts $(b',c')$ of weights $(1,0)$ and commuting ghosts $(\beta',\gamma')$ of weights $({1 \over 2},{1 \over 2})$, with central charges $-2$ and $-1$. Together with the ghosts of the $N=1$ string, whose central charge is $-15$, this gives $c_{\mathrm{gh}}=-18$, which is cancelled by twelve bosons and twelve fermions. For the BRST charge to be nilpotent, the current $J$ must have no central term in its operator product with itself, which requires
\bea
\label{eq:vnull}
v^2=0~.
\eea
This is the origin of the null vector. Physical states must be annihilated by the non-negative modes of $J$ and $\Psi$, in addition to the usual super-Virasoro constraints. As for a light-cone gauge condition, this removes the oscillators along two of the twelve directions, one timelike and one spacelike, so that the counting of states is that of a ten-dimensional string. The zero mode of $J$ imposes $v\cdot K=0$ on the momentum, and it is argued in \cite{Ooguri:1991ie} that momenta differing by a multiple of $v$ are to be identified. We will return to this last point in Section \ref{sec:nullghosts}.

The null vector $v$ cannot lie entirely within the internal space $\R^8$, since the latter is Euclidean.\footnote{One could instead take $v$ to be complex, for instance $v=e_1+ie_2$ with $e_1$ and $e_2$ orthonormal vectors of the internal space, so that $v\cdot v=0$. But in that case the condition $v\cdot K=0$ sets two components of the internal momentum to zero, and the resulting theory is not modular invariant \cite{Kutasov:1996vh}.} There are then two possibilities \cite{Ooguri:1991ie,Kutasov:1996vh}. If $v$ lies entirely in the base $\R^{2,2}$, the momenta are restricted to a plane of signature $(1,1)$, and the theory is effectively $1+1$ dimensional. If $v$ has a timelike component in the base and a spacelike component in the internal space, the momenta are restricted to a space of signature $(2,1)$, and the theory is $2+1$ dimensional. In either case, for the $E_8$ lattice the physical spectrum consists of eight bosons and eight fermions, all massless, and the theory has sixteen supercharges \cite{Kutasov:1996vh}.

The proposal of Kutasov and Martinec is that these eight bosons are the transverse coordinates of a string or of a membrane, whose worldvolume is the $1+1$ or $2+1$ dimensional target space of the $N=(2,1)$ string, and that the exact classical action for them, obtained by first relaxing the null reduction and working in $2+2$ dimensions, is
\bea
\label{eq:kmaction}
S=\int d^4x\,\sqrt{\det\big[\eta_{a\bar b}+\pd_a A_{\bar b}-\pd_{\bar b}A_a+\delta_{IJ}\,\pd_a\phi^I\pd_{\bar b}\phi^J\big]}~,
\eea
where $\phi^I$ are the eight scalars and $A$ is a vector potential which arises when $v$ is not imposed; see (3.19) of \cite{Kutasov:1996vh}. Upon null reduction to two dimensions this becomes the Nambu-Goto action in static gauge. We will meet this action again in Section \ref{sec:neutral}.

The vacua of the uncompactified $N=(2,1)$ string with Euclidean internal space were classified in \cite{Rabinovici:1997hd}. Assuming modular invariance, there are three, with massless spectra consisting of 24 chiral fermions, of 24 bosons, or of eight bosons and eight fermions, the last being the one based on the $E_8$ lattice. In this paper we will relax the assumption that the internal space is Euclidean, at the cost of restricting to tree level. An extension to higher genus requires a prescription for indefinite lattice sums, which we do not attempt here.

\subsection{Summary}
\label{sec:reviewsummary}

For ease of reference, let us now summarize the basic facts about $N=2$ strings which will be used below,
\begin{enumerate}
\item The right-movers of an $N=2$ heterotic string are those of the $N=2$ string on the base $\R^{2,2}$. At nonzero momentum they contribute a single physical state, with no oscillator excitations \cite{Ooguri:1991fp,Junemann:1997rx}. Since the right-movers carry momentum only in the base, their mass-shell condition is then $k^2=0$, and so every physical state of the heterotic string has $k^2=0$ in the base. In three-point amplitudes the right-movers contribute the factor $k_1Ik_2$, which is nonzero on one of the two kinematic branches to be described in Section \ref{sec:kinematics}.\footnote{This is the factor in the sector of vanishing instanton number of the worldsheet gauge field, which is the sector used for the three-point amplitudes of this paper. In the sectors of instanton number $\pm1$ it is replaced by invariants built from the holomorphic or from the antiholomorphic components of the momenta alone \cite{Lechtenfeld:1995zr}, as are the quantities $X$ and $\bar X$ of Section \ref{sec:kinematics}.}

\item The left-movers of the $N=(2,1)$ string are those of an $N=1$ string in twelve dimensions with a gauged null supercurrent (\ref{eq:nullcurrent}). Writing the left-moving momentum as $K=(k;p)$, with $p$ the momentum in the internal directions, the left-moving mass-shell condition in the Neveu-Schwarz sector is ${1 \over 2}K^2+N-{1 \over 2}=0$, with $N$ the oscillator level. Since $k^2=0$, the level is fixed by the internal momentum,
\bea
\label{eq:levelmatching}
n\equiv N-{1 \over 2}=-{p^2 \over 2}~.
\eea
For a Euclidean internal lattice only $p=0$ and $n=0$ are possible, which is why the standard $N=(2,1)$ string has finitely many states.
\item Amplitudes factorize into left- and right-moving parts. In particular a three-point amplitude is $A_3=(k_1Ik_2)\,C_L(1,2,3)$, with $C_L$ the three-point coupling of the left-moving $N=1$ string.
\end{enumerate}

\section{Null gauging as a cohomological reduction}
\label{sec:mechanism}

The action of the null-current gauging on the oscillator states admits a useful interpretation as an orthosymplectic cohomological reduction. We first explain the finite-dimensional construction, then identify it in the worldsheet fields. The treatment of zero modes requires an additional choice, which we describe. 

\subsection{Warm-up: cohomological reduction in finite dimensions}
\label{sec:warmup}

We begin by recalling Parisi-Sourlas dimensional reduction \cite{Parisi:1979ka}, which relates suitable superspace integrals and invariant correlation functions in $n$ even and two odd coordinates to quantities in $n-2$ even dimensions. Its simplest instance, for a sufficiently rapidly decaying function $f$, is the identity
\bea
\label{eq:parisisourlas}
\int d^nx\,d\theta^+d\theta^-\,f\big(x^2+2\,\theta^+\theta^-\big)\ \propto\ \int d^{n-2}x\,f\big(x^2\big)~,
\eea
which the reader may easily verify by expanding in the odd coordinates. The mechanism behind it is the following. The superspace admits an odd symmetry $Q$ with $Q^2=0$, which mixes one pair of even coordinates with the pair of odd ones, and the integrand is invariant under it. Functions which are $Q$-invariant and which differ by something of the form $Q(\dots)$ have the same integral, and the classes of such functions can be identified with functions on the subspace left fixed by $Q$, which is $\R^{n-2}$.

This is the prototype of a general phenomenon. Suppose a theory has a Lie superalgebra of symmetries, and let $Q$ be an odd element which squares to zero. Then $Q$ acts as a differential on any space on which the algebra acts, and one may restrict attention to the cohomology of $Q$, that is to objects annihilated by $Q$, modulo objects of the form $Q(\dots)$. In favorable circumstances the result is a smaller theory, whose symmetry algebra is the cohomology of $Q$ acting on the original algebra. This procedure was studied systematically for sigma models with supergroup symmetry in \cite{Candu:2010yg}, where it was named \emph{cohomological reduction}. A more familiar example is the BRST quantization of gauge theories, in which the two unphysical polarizations of the gauge field and the two ghosts form a ``quartet'' which drops out of the cohomology \cite{Kugo:1979gm}.

Moving on,  the case of interest to us is the following. Let $V=\R^{r,s|2}$ be the superspace with $r$ spacelike and $s$ timelike even directions, with metric $\eta$, and two odd directions with basis $e_+,e_-$, paired by a symplectic form. The linear transformations preserving both structures form the orthosymplectic supergroup, with Lie superalgebra $\osp(r,s|2)$. Its even part is $\so(r,s)\oplus\mathfrak{sp}(2,\R)$, and its odd part consists of the transformations mixing even and odd directions, which are labelled by a pair of vectors of $\R^{r,s}$ \cite{Frappat:1996pb}. To a single vector $v$ we may associate the odd element $Q_v$ acting as
\bea
\label{eq:qvmodel}
Q_v\,e_+=v~, \hspace{0.4 in} Q_v\,x=(v\!\cdot\!x)\,e_-~, \hspace{0.4 in} Q_v\,e_-=0~,
\eea
for $x$ an even vector. Applying $Q_v$ twice gives $Q_v^2\,e_+=(v\!\cdot\!v)\,e_-$ and zero on everything else, so that
\bea
\label{eq:qvsquare}
Q_v^2=0 \hspace{0.4 in} \Longleftrightarrow \hspace{0.4 in} v^2=0~.
\eea
Let us then take $v$ to be null and compute the cohomology. The kernel of $Q_v$ consists of $e_-$ and of the even vectors orthogonal to $v$; its image consists of $v$ itself and of $e_-$. The cohomology is therefore
\bea
\label{eq:cohomologymodel}
H_{Q_v}\big(\R^{r,s|2}\big)\,=\,v^\perp/v \,\,\cong\,\,\R^{r-1,s-1}~.
\eea
In words, the null vector $v$, a second vector $\bar v$ with $v\cdot\bar v\neq0$, and the two odd directions form a quartet, $e_+\to v$ and $\bar v\to e_-$, and drop out. What remains is an ordinary space with one fewer spacelike and one fewer timelike direction. Likewise, the cohomology of $Q_v$ acting on the algebra $\osp(r,s|2)$ by commutation is $\so(r-1,s-1)$. For example, both $\osp(10,2|2)$ and $\osp(6,6|2)$ have 69 even generators and 24 odd ones, and in both cases the cohomology has dimension $45$ and is purely even. It is $\so(9,1)$ in the first case and $\so(5,5)$ in the second.

The lesson is that the pair consisting of a superspace with two odd directions and a null vector is a redundant description of a space with two fewer dimensions. We will now see that this is exactly what the null current of the $N=(2,1)$ string provides.

\subsection{The null current and its ghosts}
\label{sec:nullghosts}

Recall from Section \ref{sec:heteroticreview} that the left-movers of the $N=(2,1)$ string consist of twelve bosons $X^M$ and fermions $\psi^M$, with a metric of signature $(10,2)$, together with the ghosts $(b',c')$ and $(\beta',\gamma')$ of the null supercurrent. The anticommuting ghosts have weights $(1,0)$. A weight-zero anticommuting field is nothing but an odd coordinate, and $b'$ can be written as the derivative of a second one,
\bea
\label{eq:oddcoordinates}
c'=\theta^+~, \hspace{0.4 in} b'=\pd\theta^-~,
\eea
in the same way that $\pd X^M$ is a weight-one field built from the weight-zero field $X^M$.\footnote{This is the system known as symplectic fermions \cite{Kausch:2000fu}. The fields $(b',c')$ and $(\pd\theta^-,\theta^+)$ have the same local operator products, but differ in their zero modes. In particular, $\theta^-$ has a zero mode which $b'$ does not. This difference is discussed in Appendix \ref{app:zeromodes}.} Similarly, the commuting ghosts $(\beta',\gamma')$ have weight ${1 \over 2}$, like the fermions $\psi^M$, and are the worldsheet superpartners of $\theta^\mp$. In other words, the fields
\bea
\label{eq:superspacefields}
\big(X^M\,\big|\,\theta^+,\theta^-\big) \hspace{0.4 in} \text{and} \hspace{0.4 in} \big(\psi^M\,\big|\,\gamma',\beta'\big)
\eea
are the coordinates of the superspace $\R^{10,2|2}$ and their worldsheet superpartners. The stress tensor and supercurrent of matter and null ghosts together,
\bea
\label{eq:tgextended}
T&=&-{1 \over 2}\,\pd X\!\cdot\!\pd X-{1 \over 2}\,\psi\!\cdot\!\pd\psi-b'\pd c'+{1 \over 2}\big(\beta'\pd\gamma'-\pd\beta'\,\gamma'\big)~, \nonumber\\ G&=&i\big[\psi\!\cdot\!\pd X+\beta'\pd c'-\gamma'b'\big]~,
\eea
are those of a free $N=1$ sigma model on this superspace. The central charge is $18-3=15$, the same as for ten flat dimensions, in accordance with the rule that an odd direction counts as minus one even direction.

Next consider the BRST charge of the null supercurrent $Q_v=\oint(c'\,v\cdot\pd X+\gamma'\,v\cdot\psi)$. It acts on the fields as
\bea
\label{eq:qvfields}
Q_vX^M=-v^M\,c'~, \hspace{0.3 in} Q_vb'=v\!\cdot\!\pd X~, \hspace{0.3 in} Q_v\psi^M=v^M\,\gamma'~, \hspace{0.3 in} Q_v\beta'=-v\!\cdot\!\psi~,
\eea
and annihilates $c'$ and $\gamma'$. Comparing with (\ref{eq:qvmodel}), we see that $Q_v$ acts on each of the two multiplets (\ref{eq:superspacefields}) precisely as the odd element of $\osp(10,2|2)$ acts on the coordinate functions of the superspace, up to signs. In particular, it maps the even coordinates to $v^M$ times an odd one, and an odd coordinate to the component of the even ones along $v$. It is easy to verify that
\bea
\label{eq:qvinvariance}
Q_vT=Q_vG=0 \hspace{0.1 in} \text{for all } v~, \hspace{0.4 in} Q_v^2\,b'=-v^2\,\pd c'~, \hspace{0.4 in} Q_v^2\,\beta'=-v^2\,\gamma'~,
\eea
so that $Q_v$ is a symmetry of the sigma model on $\R^{10,2|2}$ for any $v$, and is nilpotent precisely when $v$ is null, as in (\ref{eq:qvsquare}).

We can now state the main point. The full left-moving BRST charge of the $N=(2,1)$ string is the sum of two anticommuting pieces: the ordinary BRST charge of the $N=1$ string, built from the $T$ and $G$ of (\ref{eq:tgextended}), and $Q_v$. In other words,
\begin{quote}
\emph{the null-current gauging acts on the left-moving oscillator states by the orthosymplectic cohomological reduction of the free $N=1$ system on $\R^{10,2|2}$.}
\end{quote}
The outcome of the reduction is, by (\ref{eq:cohomologymodel}), a string on $v^\perp/v=\R^{9,1}$, the familiar statement that the null current removes two of the twelve directions. From this point of view the null vector is not an additional piece of data, but part of the choice of a nilpotent odd symmetry of a larger and more symmetric system.  Of course, the full string is less symmetric than its left-moving oscillators, since the base $\R^{2,2}$ is singled out by the right-movers, and the internal directions are compactified on a lattice. The position of $v$ relative to these is physically meaningful, and distinguishes the different vacua.

\subsubsection{Zero modes} The statement that the reduction removes two directions requires one qualification. For the zero modes, the constraint $v\cdot K=0$ restricts the momentum to $v^\perp$, but does not by itself identify momenta which differ by a multiple of $v$. It is this additional identification which is needed to obtain $v^\perp/v$. The standard null-gauged strings implement it globally---in Appendix A of \cite{Kutasov:1996vh}, integration over the gauge-field zero mode imposes $v\cdot K=0$, and spectral flow in the null $U(1)$ identifies the remaining null momentum labels. For a primitive null vector of a purely left-moving lattice the identification takes a simpler form. The operator $e^{iv\cdot X}$ has weight zero and a $Q_v$-exact derivative, and one can impose the additional quotient $e^{iv\cdot X}\sim1$, which identifies $p\sim p+v$ and leaves the lattice $v^\perp/v$. We refer to this as the lattice zero-mode quotient. We impose it throughout, except in Section \ref{sec:worldline}, where the null momentum label is retained before the identification. One may also ask whether the identification can be obtained from the cohomology of $Q_v$ itself, by including the zero mode of $\theta^-$. As explained in Appendix \ref{app:zeromodes}, this is possible locally, but is obstructed by the right-movers whenever $v$ has a component in the base. In any case, the $(5,5)$ string requires no null gauging and therefore avoids all of these issues.

\subsection{Real forms and an invariant}
\label{sec:realforms}

Nothing in the above depended on the signature $(10,2)$. Let us then consider the following family of left-moving theories: an $N=1$ string on
\bea
\label{eq:family}
\R^{2,2}\times\II_{a,b}~,
\eea
with $g$ mutually orthogonal null supercurrents gauged. Here $\II_{a,b}$ is an even self-dual lattice of signature $(a,b)$, on which the purely left-moving directions are compactified. Such lattices exist if and only if $a-b$ is a multiple of eight. Each null supercurrent reduces the critical dimension count by two, so criticality requires $a+b=6+2g$. The standard $N=(2,1)$ string is $(a,b;g)=(8,0;1)$.

\begin{table}[t]
\centering
\begin{tabular}{@{}lccc@{}}
\toprule
 & $\sigma=8$ & $\sigma=0$ & $\sigma=-8$\\
\midrule
Reduction & $\R^{10,2|2}\to\R^{9,1}$ & $\R^{6,6|2}\to\R^{5,5}$ & $\R^{2,10|2}\to\R^{1,9}$\\
Minimal lattice and $\#$ of null currents & $E_8$, $g=1$ & $\II_{3,3}$, $g=0$ & $E_8(-1)$, $g=1$\\
Transverse directions of a string & $(8,0)$ & $(4,4)$ & $(0,8)$\\
\bottomrule
\end{tabular}
\caption{The three real forms of the reduction for the $N=(2,1)$ string. The middle column is the only one for which no null current is needed.}
\label{tab:realforms}
\end{table}

Given such a theory, we may add a hyperbolic plane $U=\II_{1,1}$ and gauge an additional null supercurrent along a primitive null vector of $U$. The added oscillators form a quartet. With the lattice zero-mode quotient mentioned above, the added lattice directions are removed as well. In this sense the move
\bea
\label{eq:move}
(a,b;g)\to(a+1,b+1;g+1)
\eea
is an equivalence, and the combination
\bea
\label{eq:sigma}
\sigma=a-b\in8\Z
\eea
is invariant under it. It is the signature of the lattice which remains once everything that the null currents cancel has been stripped away.\footnote{We stress that $\sigma$ is an invariant, not a classification. Theories with the same $\sigma$ can differ in the choice of lattice, null vectors, and GSO projection.} The $g$ gauged null vectors are linearly independent and mutually orthogonal, so they span an isotropic subspace of the left-moving space $\R^{2,2}\oplus(\II_{a,b}\otimes\R)$. Since this space has signature $(a+2,b+2)$, such a subspace has dimension at most $\min(a+2,b+2)$, and hence $g\leq\min(a+2,b+2)$. Together with $a+b=6+2g$, this gives $|a-b|\leq10$, and even unimodularity then leaves $\sigma=8,0,-8$. These are the three real forms in (\ref{eq:introrealforms}). They are listed in Table \ref{tab:realforms}.

The crucial difference between the three cases is the following. For $\sigma=\pm8$ the minimal representative has a definite lattice, $E_8$ or its negative, which contains no real null vectors. At least one null vector must then have a component in the base, so the null gauging cannot be eliminated by reducing the internal lattice alone. 
For $\sigma=0$, on the other hand, the lattices are $\II_{g+3,g+3}$, which are indefinite, and \emph{all} of the null vectors can be placed inside the lattice. Applying the move (\ref{eq:move}) in reverse, we arrive at $g=0$, a theory with no null current at all. This is the $(5,5)$ string, to which we turn in the next section.

\section{The $(5,5)$ string}
\label{sec:model}

We now turn to the theory singled out at the end of the previous section. We first define it, and then determine its spectrum and its supersymmetry.

\subsection{Definition}
\label{sec:definition}

The $(5,5)$ string is the heterotic string whose right-movers are those of the $N=2$ string on $\R^{2,2}$, and whose left-movers are those of the $N=1$ string on
\bea
\label{eq:modeltarget}
\R^{2,2}\times\II_{3,3}~,
\eea
with the usual GSO projection, which is recalled in Section \ref{sec:susy}, and with no null current. Concretely, the left-moving fields are the four bosons $X^\mu$ of the base, shared with the right-movers; six chiral bosons $Y^I$, $I=1,\dots,6$, with a metric $\eta_{IJ}$ of signature $(3,3)$, whose momenta $p_I$ take values in the lattice $\II_{3,3}$; and ten fermions $\psi^M$, $M=(\mu,I)$, with a metric of signature $(5,5)$. The left-moving constraints are the ordinary $N=1$ super-Virasoro constraints, and the theory is critical since it has ten bosons and ten fermions. At the oscillator level it is related to the theory on $\R^{2,2}\times\II_{4,4}$, with total signature $(6,6)$, by gauging a null supercurrent inside the lattice. Strings of $N=(2,1)$ type in spacetimes of signature $(n,n)$ have been considered before \cite{Rudychev:1997ui}; in that proposal it is the base shared by left- and right-movers which is enlarged, and several null currents are needed. Here the base remains $\R^{2,2}$.

A few words on the lattice are in order. $\II_{3,3}$ is the unique even self-dual lattice of signature $(3,3)$. It is the sum of three copies of the hyperbolic plane $U$, and so has a basis of null vectors $\ell_j,\ell'_j$, $j=1,2,3$, with
\bea
\label{eq:latticebasis}
\ell_j\!\cdot\!\ell_k=\ell'_j\!\cdot\!\ell'_k=0~, \hspace{0.4 in} \ell_j\!\cdot\!\ell'_k=\delta_{jk}~.
\eea
A second useful description is as the lattice $\wedge^2\Z^4$ of integral two-forms in four dimensions, with the inner product $p\cdot p'$ defined by $p\wedge p'=(p\cdot p')\,dx^1dx^2dx^3dx^4$. This realizes the double cover $\mathrm{SL}(4,\R)\to \mathrm{SO}_0(3,3)$, which will play an important role below. In this description a lattice vector is null precisely when it is decomposable, $p=a\wedge b$.

It should be stressed that three of the six chiral bosons $Y^I$ are timelike. Together with the two times of the base this gives the left-movers five timelike directions, only one of which is removed by the $N=1$ constraints. We will see the consequences of this shortly.

\subsection{Spectrum}
\label{sec:spectrum}

By (\ref{eq:levelmatching}), every physical state has null momentum $k$ in the base, and a state with internal momentum $p$ sits at left-moving level $n=-p^2/2$. We will refer to $p$ as the charge of the state. It is the eigenvalue of the zero modes of the six currents $i\pd Y^I$, and from the point of view of the base it is a conserved and quantized label, just as the momentum along a compact direction is a charge in Kaluza-Klein theory. Since the lattice is indefinite, all levels $n=0,1,2,\dots$ occur. Let us describe the first two.

\paragraph{Level zero: neutral states.} For $p=0$ the left-moving ten-momentum is $K=(k;0)$ and the states are the massless states of the ten-dimensional $N=1$ string. In the Neveu-Schwarz sector they are $\varepsilon\cdot\psi_{-1/2}|K\rangle$, with $\varepsilon\cdot K=0$ and $\varepsilon\sim\varepsilon+K$, which leaves eight polarizations: two along the base, which we will call gluons, and six scalars along the internal directions. In the Ramond sector there are eight states, described by a spinor $u$ obeying the Dirac equation $\slashed{K}u=0$. From the four-dimensional point of view the eight fermions carry an index in the fundamental $4$ or its dual $4^*$ of $\mathrm{SL}(4,\R)$, according to their helicity.

Multiplying by the single state of the right-movers, we find that the neutral sector of the $(5,5)$ string consists of $8+8$ states. The left-moving states have helicities $1,{1 \over 2},0,-{1 \over 2},-1$, as in $N=4$ Yang-Mills theory. In the description developed in Section \ref{sec:neutral}, the right-movers shift each of these by one unit, so that the fields have
\bea
\label{eq:helicities}
\text{helicity:} \hspace{0.1 in} 2~,\ \ {3 \over 2}~,\ \ 1~,\ \ {1 \over 2}~,\ \ 0 \hspace{0.4 in} \text{multiplicity:} \hspace{0.1 in} 1~,\ \ 4~,\ \ 6~,\ \ 4~,\ \ 1~.
\eea
This is the positive-helicity half of the multiplet of $N=4$ supergravity. Note that the inner product on these states is not positive. Indeed, of the eight bosons, four have negative norm, namely one of the two gluon polarizations and the three scalars along the timelike directions of the lattice.

\paragraph{Level zero: null charges.} For $p\neq0$ with $p^2=0$ the level is again zero, and the left-moving states are again the massless states of the ten-dimensional string, now with ten-momentum $K=(k;p)$. There are thus 16 states for every null vector of the lattice. Since the right-movers do not see the internal charge, these states are massless in the base for any $p$.

\paragraph{Level one.} For $p^2=-2$ the left-moving states are those of the first massive level of the ten-dimensional open superstring, with $K^2=-2$. In the Neveu-Schwarz sector they are built from $\psi_{-3/2}$, $\alpha_{-1}\psi_{-1/2}$, and $\psi_{-1/2}\psi_{-1/2}\psi_{-1/2}$. After imposing the constraints and quotienting by spurious states one finds 128 bosons, and likewise 128 fermions from the Ramond sector. In signature $(5,5)$ the 128 bosons form the representations $44+84$ of the little group $\mathrm{SO}(5,4)$, and their norms have signature $(64,64)$. With respect to the base these states are massless, and the left-moving factor carries helicities from $-2$ to $2$, with multiplicities $1,28,70,28,1$ for the bosons.

\subsection{The Ramond sector and supersymmetry}
\label{sec:susy}

The fermionic states are created by spin fields, and since the signature is unusual it is worth checking that the standard construction goes through. Let us group the ten fermions into five pairs, each consisting of one spacelike and one timelike direction, and bosonize each pair, $\psi^\pm_i=(\psi^{s_i}\pm\psi^{t_i})/\sqrt2=e^{\pm\phi_i}$. The spin fields are $S_\epsilon=e^{\epsilon\cdot\phi/2}$ with $\epsilon_i=\pm1$. Each pair spans a plane of signature $(1,1)$, so the Cartan generator which acts on it is a boost rather than a rotation. A boost has real eigenvalues and real eigenvectors, namely the null combinations $\psi^\pm_i$, so no factors of $i$ appear in the bosonization. In signature $(5,5)$ all ten directions can be paired in this way, with the consequence that the bosons $\phi_i$, the weights, and the spin fields are all real.

Note that the mutual locality of the vertex operators depends only on their weights, and not on the signature, so the GSO projection is the usual one, with an odd number of fermions in the Neveu-Schwarz sector, and spin fields of one chirality in the Ramond sector. In particular, the Neveu-Schwarz ground state, which would be a tachyon, is removed. The sixteen surviving spin fields, which we denote by $S_\alpha$, form a Majorana-Weyl spinor of $\mathrm{Spin}(5,5)$, the Majorana condition being simply that the coefficients are real. A fermionic state is thus described by a real sixteen-component spinor wavefunction $u^\alpha$, in the same way that a bosonic state is described by a polarization vector $\varepsilon^M$. In terms of these, the three-point vertex of two fermions and a boson is proportional to $u_1^TC\Gamma_Mu_2\,\varepsilon_3^M$, where $u_1$ and $u_2$ are the wavefunctions of the two fermions, $\varepsilon_3$ is the polarization of the boson, $\Gamma_M$ are the gamma matrices of $\mathrm{Spin}(5,5)$, and $C$ is the charge conjugation matrix, defined by the property that $u_1^TC\Gamma_Mu_2$ transforms as a vector. This vertex is the unique one allowed by covariance, since the vector representation appears exactly once in the symmetric product of two spinors of the same chirality.

The supercharges are the zero modes of the spin fields in the $-{1 \over 2}$ picture, $Q_\alpha=\oint e^{-\varphi/2}S_\alpha$, with $\varphi$ the bosonized superghost, and from the operator product of two spin fields one finds
\bea
\label{eq:susyalgebra}
\{Q_\alpha,Q_\beta\}=(C\Gamma^M)_{\alpha\beta}\,K_M~, \hspace{0.4 in} K_M=(k_\mu;p_I)~.
\eea
This is the ten-dimensional supersymmetry algebra, but with an important difference in interpretation: the components of $K$ along the internal directions are the lattice charges. To see what this means in four-dimensional terms, decompose the sixteen supercharges under $\mathrm{Spin}(2,2)\times \mathrm{Spin}(3,3)\cong \mathrm{SL}(2,\R)\times \mathrm{SL}(2,\R)\times \mathrm{SL}(4,\R)$ as $(2,1,4)+(1,2,4^*)$, and write them as $Q_{\alpha A}$ and $\tilde Q_{\dot\alpha}{}^A$. Then (\ref{eq:susyalgebra}) becomes
\bea
\label{eq:susysplit}
\{Q_{\alpha A},\tilde Q_{\dot\alpha}{}^B\}=\delta_A{}^B\,k_{\alpha\dot\alpha}~, \hspace{0.4 in} \{Q_{\alpha A},Q_{\beta B}\}=\epsilon_{\alpha\beta}\,Z_{AB}~, \hspace{0.4 in} Z_{AB}=p_I\,\Sigma^I_{AB}~,
\eea
together with a similar relation for $\tilde Q$, where $\Sigma^I_{AB}$ are six antisymmetric $4\times4$ matrices which convert a vector $p_I$ of $\mathrm{SO}(3,3)$ into an antisymmetric tensor $Z_{AB}$ of $\mathrm{SL}(4,\R)$, in the same way that the Pauli matrices convert a four-vector into a bispinor. In the description of the lattice as $\wedge^2\Z^4$ given in Section \ref{sec:definition}, $Z_{AB}$ is simply the two-form which represents $p$. This is the algebra of $N=4$ supersymmetry in $2+2$ dimensions, with the six lattice charges appearing as central charges.

\subsection{Multiplets}
\label{sec:multiplets}

The representations of (\ref{eq:susyalgebra}) are determined by the rank of the symmetric form $(C\Gamma^M)K_M$ on the sixteen supercharges. For nonzero $K$ with $K^2=0$, half of the supercharges annihilate the states, and the multiplets have $2^4=16$ states, as realized by the level-zero spectrum. For $K^2\neq0$ the rank is sixteen and the multiplets have $2^8=256$ states. Since $K^2=p^2$, the size of a multiplet is determined by the norm of its charge, as summarized in Table \ref{tab:multiplets}.

\begin{table}[t]
\centering
\begin{tabular}{@{}llll@{}}
\toprule
Charge & States & Multiplicities by helicity & Unbroken supercharges\\
\midrule
$p=0$ & 16 & $1,4,6,4,1$ & 8\\
$p\neq0$, $p^2=0$ & 16 & $1,4,6,4,1$ & 8\\
$p^2=-2n<0$ & 256 & $1,8,28,56,70,56,28,8,1$ & 0\\
\bottomrule
\end{tabular}
\caption{Multiplets of the $(5,5)$ string. The multiplicities are those of the helicities of the left-moving factor, $1,{1 \over 2},\dots,-1$ in the first two rows and $2,{3 \over 2},\dots,-2$ in the last for the first timelike level. For a given charge of level $n=1,2,3,4,\dots$ the number of such multiplets is $1,9,60,329,\dots$}
\label{tab:multiplets}
\end{table}

Two features of this table deserve comment. First, the long multiplets are massless in the base, even though they sit at a massive level of the left-moving string. In an ordinary string, a state of level $n$ satisfies the mass-shell condition ${1 \over 2}K^2+n=0$ by acquiring a mass, $k^2=-2n$. Here the right-movers require $k^2=0$, and the condition is satisfied instead by the internal charge, $p^2=-2n$, which is possible because the lattice has timelike directions. The states of level $n$ reach helicity $n+1$ in the left-moving factor. The charged sector of the $(5,5)$ string is thus a theory of massless fields of arbitrarily high helicity. Second, there is no BPS bound relating charges and masses, as there would be in a unitary theory, since the form $(C\Gamma^M)K_M$ is indefinite. Indeed, it has signature $(4,4)$ on its image for null $K$ and $(8,8)$ for timelike $K$.

\subsection{Cautionary remarks}
\label{sec:scope}

We have seen that the $(5,5)$ string has negative-norm states already in its neutral sector, as a direct consequence of the timelike directions in the lattice. There is a second, related difficulty. The lattice $\II_{3,3}$ contains infinitely many vectors of each norm $-2n$, so every left-moving level contains infinitely many states, and the sum over the lattice which would appear in the torus amplitude does not converge. The ordinary torus lattice sum is therefore undefined without an additional prescription. Both features are shared with the lattice vertex algebras of indefinite lattices familiar from the study of generalized Kac-Moody algebras, of which the left-moving sector of the $(5,5)$ string is an example.

For these reasons we study the theory locally and at tree level, using its vertex operators, its BRST cohomology, and its sphere amplitudes, all of which are perfectly well-behaved. This is sufficient for the questions we wish to ask, which concern the classical target-space theory. Note that the lattice charge is conserved, so that sectors closed under addition of charges are closed under tree-level interactions. In particular the neutral sector is by itself a consistent classical theory, with finitely many fields, and we begin with it.

\section{The neutral sector and its cubic superfield action}
\label{sec:neutral}

In this section we construct a cubic target-space theory for the sixteen neutral states. We begin with the three-point amplitudes, then discuss the bosonic action and its field equations, and finally show that the complete theory, fermions included, is described by a single superfield.

\subsection{Kinematics in $2+2$ dimensions}
\label{sec:kinematics}

Before computing the amplitudes, let us review some standard computational tools that may be unfamiliar to the reader. Since all states of the theories we consider are massless in $\R^{2,2}$, it is convenient to use spinor variables \cite{Elvang:2015rqa}, which in signature $(2,2)$ are real. The Lorentz group is locally $\mathrm{SL}(2,\R)\times \mathrm{SL}(2,\R)$, a vector is a bispinor $k_{\alpha\dot\alpha}$, and a null vector is one of rank one,
\bea
\label{eq:spinors}
k_{\alpha\dot\alpha}=\lambda_\alpha\tilde\lambda_{\dot\alpha}~, \hspace{0.4 in} 2\,k_i\!\cdot\!k_j=\langle ij\rangle[ij]~,
\eea
with $\langle ij\rangle:=\epsilon^{\alpha\beta}\lambda_{i\alpha}\lambda_{j\beta}$ and $[ij]:=\epsilon^{\dot\alpha\dot\beta}\tilde\lambda_{i\dot\alpha}\tilde\lambda_{j\dot\beta}$. Our conventions are given in Appendix \ref{app:conventions}. In terms of these variables, we then have the following results.

\paragraph{Three-point kinematics.} Three null momenta which add up to zero satisfy $k_i\cdot k_j=0$, so for each pair either $\langle ij\rangle$ or $[ij]$ vanishes. It follows that for three-point amplitudes there are two branches: either all three $\lambda_i$ are proportional, or all three $\tilde\lambda_i$ are. In Minkowski signature these configurations exist only for complex momenta, but in signature $(2,2)$ they are real, and three-point amplitudes are honest observables.

\paragraph{The complex structure.} A complex structure on $\R^{2,2}$ compatible with the metric is a rotation generator in one of the two $\mathrm{SL}(2,\R)$ factors. We take it to act on the undotted index, $(Ik)_{\alpha\dot\alpha}=I_\alpha{}^\beta k_{\beta\dot\alpha}$ with $I^2=-1$. The three-point factor of the $N=2$ string is then
\bea
\label{eq:rightfactor}
k_1Ik_2\propto\epsilon^{\alpha\beta}\,(I\lambda_1)_\alpha\lambda_{2\beta}\,[12]~,
\eea
which vanishes on the branch where the $\tilde\lambda_i$ are proportional. On the other branch, writing $\lambda_i=t_i\lambda$, we have
\bea
\label{eq:rightfactorbranch}
k_1Ik_2=t_1t_2\,[12]\,Q(\lambda)~, \hspace{0.4 in} Q(\lambda)={1 \over 2}\big(\lambda_0^2+\lambda_1^2\big)~,
\eea
which is nonzero for generic real momenta. In other words, the $N=2$ string has amplitudes on one of the two branches only. This is the sense in which it is self-dual---on the surviving branch, linearized field strengths built from the momenta are self-dual.

\paragraph{Holomorphic components.} The eigenvectors of $I$ with eigenvalue $-i$ form a plane $\Lambda$ in the complexification of $\R^{2,2}$, consisting of the vectors $\mu_\alpha\tilde\lambda_{\dot\alpha}$ with $\mu=(1,i)$ fixed and $\tilde\lambda$ arbitrary. Its complex conjugate $\bar\Lambda$ is the other eigenspace. Both planes are totally null, i.e. any two vectors in the same plane are orthogonal, and in particular every vector in it is null. In complex coordinates $z^a$, $a=1,2$, adapted to $I$, a plane wave is $e^{i(k_az^a+k_{\bar a}\bz^{\bar a})}$, where the holomorphic components $k_a$ are the components of $k$ along $\Lambda$, and for real momenta $k_{\bar a}$ is the complex conjugate of $k_a$. The basic invariants of two momenta are
\bea
\label{eq:xdef}
X(1,2)=\epsilon^{ab}\,k_{1a}k_{2b}~, \hspace{0.4 in} \bar X(1,2)=\epsilon^{\bar a\bar b}\,k_{1\bar a}k_{2\bar b}~.
\eea
For three real null momenta on the branch (\ref{eq:rightfactorbranch}), one finds
\bea
\label{eq:xxbar}
k_1Ik_2=4\,e^{i\vartheta(\lambda)}\,\bar X(1,2)~, \hspace{0.3 in} e^{i\vartheta(\lambda)}={\lambda_0-i\lambda_1 \over \lambda_0+i\lambda_1}~, \hspace{0.3 in} \big(k_1Ik_2\big)^2=16\,X(1,2)\,\bar X(1,2)~.
\eea
Thus $\bar X$ differs from the three-point factor of the $N=2$ string by a constant normalization and a phase depending on the common spinor $\lambda$. For the neutral states this phase can be absorbed in the normalization of the external states. Multiplying each state of left-moving helicity $h$ by $e^{-ih\vartheta}$ leaves the two-point function unchanged, since it pairs helicities $h$ and $-h$, and multiplies the three-point amplitudes by $e^{-i\vartheta}$, since as we will see in Section \ref{sec:threepoint} their left-moving helicities add up to $+1$. The charged states are discussed in Section \ref{sec:frame}. The reason for introducing $\bar X$ is that it is defined for arbitrary momenta, on or off shell, whereas $k_1Ik_2$ agrees with it only for three null momenta. The cubic theories of Section \ref{sec:superamplitudes} and Section \ref{sec:charged} are built with $\bar X$. For the charged states the tree amplitudes would not vanish if $k_1Ik_2$ were used instead, as discussed in Section \ref{sec:tower}.

\subsection{Three-point amplitudes}
\label{sec:threepoint}

As reviewed in Section \ref{sec:reviewsummary}, a three-point amplitude is the product of the right-moving factor $k_1Ik_2$ and the three-point coupling of the left-moving $N=1$ string. For three bosons, and for two fermions and a boson, the latter are the familiar couplings of ten-dimensional super Yang-Mills theory, so that
\bea
\label{eq:athreeboson}
A_3^{\mathrm{bbb}}&=&\big(k_1Ik_2\big)\,\big[(\varepsilon_1\!\cdot\!\varepsilon_2)(\varepsilon_3\!\cdot\!K_1)+(\varepsilon_2\!\cdot\!\varepsilon_3)(\varepsilon_1\!\cdot\!K_2)+(\varepsilon_3\!\cdot\!\varepsilon_1)(\varepsilon_2\!\cdot\!K_3)\big]~, \nonumber\\ A_3^{\mathrm{ffb}}&=&{1 \over \sqrt2}\,\big(k_1Ik_2\big)\,u_1^TC\Gamma_Mu_2\,\varepsilon_3^M~,
\eea
with polarizations and inner products those of signature $(5,5)$, and the Ramond zero-mode normalization $\psi_0^M=\Gamma^M/\sqrt2$ used in Appendix \ref{app:correlators}. To exhibit the structure of these amplitudes, it is useful to use the spinor-helicity formalism reviewed above. Recall that the amplitude is nonzero only on the branch $\lambda_i=t_i\lambda$. Let us label the sixteen states of each leg by four odd variables $\eta_i^A$ in the $4$ of $\mathrm{SL}(4,\R)$, in the standard way, where the coefficient of a monomial of degree $r$ in $\eta_i$ is the state of left-moving helicity $1-r/2$. Then we find
\bea
\label{eq:superamp}
A_3=\underbrace{t_1t_2\,[12]\,Q(\lambda)}_{\text{right-movers}}\ \times\ \underbrace{{\delta^{(4)}\big(\eta_1[23]+\eta_2[31]+\eta_3[12]\big) \over [12][23][31]}}_{\text{left-movers}}~.
\eea
The right-moving factor is $k_1Ik_2$, in the form (\ref{eq:rightfactorbranch}). The left-moving factor is the three-point superamplitude of $N=4$ super Yang-Mills theory on this kinematic branch. Equation (\ref{eq:superamp}) can be checked component-by-component. The amplitudes are nonzero precisely when the left-moving helicities sum to $+1$; the dependence on the momenta of each component is the one dictated by (\ref{eq:superamp}), for instance $[12]^3/[23][31]$ for two gluons of positive and one of negative helicity; the scalars couple through the $(3,3)$ metric and the fermions through the matrices $\Sigma^I_{AB}$ of (\ref{eq:susysplit}); and the supersymmetry Ward identities relating the bosonic and fermionic amplitudes hold for all sixteen supercharges.

The $(5,5)$ string is thus a ``double copy'' in the simplest sense. Its three-point amplitudes are those of $N=4$ Yang-Mills theory, with the color factor replaced by the structure constant $k_1Ik_2$ of the Poisson algebra (\ref{eq:poisson}). This is the structure expected of self-dual $N=4$ supergravity.

\subsection{The bosonic action and its field equations}
\label{sec:bosonicaction}

For the bosons, an action reproducing these amplitudes is known. It is the determinant action (\ref{eq:kmaction}) of Kutasov and Martinec, the only changes being that the scalars are now six in number with a metric $\eta_{IJ}$ of signature $(3,3)$, and that no null reduction is imposed,
\bea
\label{eq:action}
S=\int d^4x\,\sqrt{-\det g_{a\bar b}}~, \hspace{0.4 in} g_{a\bar b}=\eta_{a\bar b}+\pd_aA_{\bar b}-\pd_{\bar b}A_a+\eta_{IJ}\,\pd_a\phi^I\pd_{\bar b}\phi^J~.
\eea
Here $\eta_{a\bar b}=\mathrm{diag}(1,-1)$ is the flat metric in the complex coordinates of Appendix \ref{app:conventions}. Its determinant is $-1$, which is the origin of the sign under the square root. We impose $A_{\bar a}=-(A_a)^*$ and take the $\phi^I$ to be real, so that $g_{a\bar b}$ is hermitian. We assume $\det g<0$, as is the case for fields close to the flat background, so that the square root is real. The field $A$ has two physical polarizations, which together with the six scalars account for the eight bosons of Section \ref{sec:spectrum}. One can verify that the cubic vertices of (\ref{eq:action}) reproduce (\ref{eq:athreeboson}), and that for four scalars the contact term and the exchange of $A$ cancel on shell, as in \cite{Kutasov:1996vh}.

What does this action describe? To answer this, let us write its equations of motion in a more geometric form. Define
\bea
\label{eq:rhodef}
\rho=\sqrt{-\det g}~, \hspace{0.4 in} \hat g_{a\bar b}={g_{a\bar b} \over \rho}~, \hspace{0.4 in} \Delta(F,G)={1 \over 2}\,\epsilon^{ac}\epsilon^{\bar b\bar d}\,\pd_a\pd_{\bar b}F\,\pd_c\pd_{\bar d}G~,
\eea
so that $\det\hat g=-1$ and $\Delta(F,F)=\det(\pd_a\pd_{\bar b}F)$. Then the Euler-Lagrange equations of (\ref{eq:action}) are equivalent to the following three statements,
\begin{enumerate}
\item The metric $\hat g$ is K\"ahler, with a potential obeying the Pleba\'nski equation,
\bea
\label{eq:plebanski}
\hat g_{a\bar b}=\pd_a\pd_{\bar b}\Omega~, \hspace{0.4 in} \Delta(\Omega,\Omega)=-1~;
\eea
\item The scalars are harmonic with respect to $\hat g$,
\bea
\label{eq:maxwell}
\Delta(\Omega,\phi^I)=0~;
\eea
\item The conformal factor obeys a Poisson equation, sourced by the scalars,
\bea
\label{eq:dilaton}
\Delta(\Omega,\rho)=-\eta_{IJ}\,\Delta(\phi^I,\phi^J)~.
\eea
\end{enumerate}
The derivation is short. The equations for $A$ state that the matrix $\rho\,g^{-1}$ is divergence-free in both indices, which for a $2\times2$ matrix of fixed determinant is the K\"ahler condition on $\hat g$. The equations for $\phi^I$ then reduce to (\ref{eq:maxwell}). Finally, (\ref{eq:dilaton}) follows from the identity $\epsilon^{ac}\epsilon^{\bar b\bar d}\pd_c\pd_{\bar d}\,g_{a\bar b}=-2\,\eta_{IJ}\Delta(\phi^I,\phi^J)$, which holds for any $g$ of the form (\ref{eq:action}).

The interpretation is as follows. Equation (\ref{eq:plebanski}) says that $\hat g$ is a Ricci-flat K\"ahler metric, that is a solution of self-dual gravity, exactly as for the closed $N=2$ string. Equation (\ref{eq:maxwell}) says that the $\phi^I$ are potentials for six self-dual Maxwell fields $F^I_{a\bar b}=\pd_a\pd_{\bar b}\phi^I$ on this background, with the conventional factor of $i$ inserted if the curvature is written as a real two-form. And (\ref{eq:dilaton}) says that $\rho$ is a scalar sourced by these Maxwell fields. In terms of helicities, $\Omega$, $\phi^I$, and $\rho$ carry helicity $2$, $1$, and $0$, in agreement with (\ref{eq:helicities}), and the two polarizations of $A$ are $\Omega$ and $\rho$. Note that each field is sourced only by fields of higher helicity, and $\rho$ appears linearly. 

Two descriptions of this system have been previously identified in the literature, and one can easily see that they are the same system in two frames. In terms of $g=\rho\,\hat g$, which is a hermitian metric with torsion, the equations are those of the $(2,1)$ sigma model geometry of Hull \cite{Hull:1997kk}. In terms of the Ricci-flat metric $\hat g$, they have the structure of the bosonic sector of self-dual $N=4$ supergravity, as described for instance in \cite{Mason:2007ct}. What is special to the $(5,5)$ string is the indefinite metric $\eta_{IJ}$, which as we will see is exactly what is needed for a real superfield description.

\subsection{A dual cubic action}
\label{sec:dual}

The action (\ref{eq:action}) is non-polynomial, and produces contact terms at every order which must cancel against exchange diagrams if the higher-point amplitudes are to vanish, as was emphasized in \cite{Kutasov:1996vh}. However, the form of the field equations above suggests that this complexity is an artifact of the variable $A$, and that in terms of $\Omega$ and $\rho$ the theory should be cubic. 
This is indeed the case, as was shown by Abou Zeid and Hull \cite{AbouZeid:1997he}. We review their derivation in the present notation, with the six scalars included.

The key is to remove the square root by introducing auxiliary fields, in the same way that the Polyakov action removes the square root of the Nambu-Goto action. Let $N^{\bar ba}$ be an auxiliary $2\times2$ matrix and $\rho$ an auxiliary scalar, and consider the action
\bea
\label{eq:firstorder}
S'=-\int d^4x\,\Big[{1 \over 2}\,N^{\bar ba}g_{a\bar b}-{1 \over 2}\,\rho\,\big(1+\det N\big)\Big]~.
\eea
The fields $N$ and $\rho$ appear without derivatives, and may be eliminated by their equations of motion. That of $\rho$ sets $\det N=-1$, and that of $N$ then gives $N=-\rho\,g^{-1}$. Taking the determinant of this relation one finds $\rho^2=-\det g$. We take the positive root, so that $\rho$ is the conformal factor of (\ref{eq:rhodef}), and $N=-\hat g^{-1}$. Substituting back gives $S'=\int d^4x\,\sqrt{-\det g}$. The action (\ref{eq:firstorder}) is therefore classically equivalent to (\ref{eq:action}). Its advantage is that $g$, and hence the field $A$, appears linearly. The field $A$ then acts as a Lagrange multiplier imposing $\pd_aN^{\bar ba}=\pd_{\bar b}N^{\bar ba}=0$. These conditions are solved locally by $N^{\bar ba}=\epsilon^{ac}\epsilon^{\bar b\bar d}\pd_c\pd_{\bar d}\Omega$ for a real function $\Omega$. This is consistent with (\ref{eq:plebanski}). Indeed, since $\det\hat g=-1$, the inverse of $\hat g$ is $-\epsilon^{ac}\epsilon^{\bar b\bar d}\hat g_{c\bar d}$, so that on shell $N^{\bar ba}=\epsilon^{ac}\epsilon^{\bar b\bar d}\hat g_{c\bar d}$. With $N$ of this form, the term of (\ref{eq:firstorder}) involving the constant background metric is a total derivative, and we arrive at
\bea
\label{eq:dual}
S_{\mathrm{dual}}=\int d^4x\,\Big[{1 \over 2}\,\rho\,\big(1+\det(\pd_a\pd_{\bar b}\Omega)\big)-{1 \over 2}\,\eta_{IJ}\,N^{\bar ba}\,\pd_a\phi^I\pd_{\bar b}\phi^J\Big]~,
\eea
in which $N^{\bar ba}$ stands for $\epsilon^{ac}\epsilon^{\bar b\bar d}\pd_c\pd_{\bar d}\Omega$. The action (\ref{eq:dual}) is cubic. Its Euler-Lagrange equations are precisely (\ref{eq:plebanski})--(\ref{eq:dilaton}), with the multiplier $\rho$ equal on shell to the conformal factor of (\ref{eq:rhodef}). At the linearized level, the two polarizations of $A$ are mapped to $\Omega$ and to $\rho$, so that the duality exchanges a field with two polarizations for two fields with one each.

\subsection{One superfield}
\label{sec:superfield}

We now come to the main result of this section. The structure of (\ref{eq:plebanski})--(\ref{eq:dilaton}), with each helicity sourced by higher ones, is characteristic of a superfield. Let $\theta^A$, $A=1,\dots,4$, be four real odd coordinates in the $4$ of $\mathrm{SL}(4,\R)$, and consider the real scalar superfield
\bea
\label{eq:superfield}
\Phi=\Omega+\theta^A\psi_A+{1 \over 2}\,\theta^A\theta^B\phi_{AB}+{1 \over 6}\,\theta^A\theta^B\theta^C\epsilon_{ABCD}\,\chi^D+\theta^1\theta^2\theta^3\theta^4\rho~.
\eea
Its components are a scalar $\Omega$, an antisymmetric tensor $\phi_{AB}$, a scalar $\rho$, and odd fields $\psi_A$ and $\chi^A$. The antisymmetric tensor of $\mathrm{SL}(4,\R)$ is the vector of $\mathrm{SO}(3,3)$, so $\phi_{AB}$ is equivalent to six scalars $\phi^I$, with $\eta_{IJ}\phi^I\phi^J={1 \over 8}\epsilon^{ABCD}\phi_{AB}\phi_{CD}$. The components of $\Phi$ are thus $8+8$ real fields, with exactly the quantum numbers of the neutral sector. Note that it is the signature $(3,3)$ which makes this possible: a real superfield of this kind requires the scalars to form a real antisymmetric tensor of $\mathrm{SL}(4,\R)$.

We propose the following equation and action as the neutral target theory,
\bea
\label{eq:superpleb}
\det\big(\pd_a\pd_{\bar b}\Phi\big)=-1~, \hspace{0.4 in} S={1 \over 2}\int d^4x\,d^4\theta\,\Big[\Phi+{1 \over 3}\,\Phi\,\det\big(\pd_a\pd_{\bar b}\Phi\big)\Big]~.
\eea
This is the Pleba\'nski equation for a superfield. Expanding in $\theta$, define $\Box_\Omega F=2\Delta(\Omega,F)=N^{\bar ba}\pd_a\pd_{\bar b}F$. This operator is proportional to the scalar wave operator in the background $\hat g$, with $N=-\hat g^{-1}$. The component equations are then
\bea
\label{eq:components}
\Delta(\Omega,\Omega)=-1~, \hspace{0.4 in} \Box_\Omega\psi_A=0~, \hspace{0.4 in} \Box_\Omega\phi_{AB}=2\,\Delta(\psi_A,\psi_B)~,\hspace{0.7 in}
\nonumber\\
\Box_\Omega\chi^A=\epsilon^{ABCD}\Delta(\psi_B,\phi_{CD})~,  \hspace{0.4 in} \Box_\Omega\rho=-{1 \over 4}\,\epsilon^{ABCD}\Delta(\phi_{AB},\phi_{CD})-2\,\Delta(\psi_A,\chi^A)~.\,\,\,
\eea
Setting the odd fields to zero, these are exactly (\ref{eq:plebanski})--(\ref{eq:dilaton}), and the $\theta^4$ component of the Lagrangian in (\ref{eq:superpleb}) is the dual Lagrangian (\ref{eq:dual}). The superfield action is therefore the natural completion of the bosonic action by the fermions, with the component Lagrangian
\bea
\label{eq:compaction}
\mathcal{L}&=&{1 \over 2}\,\rho\,\big(1+\Delta(\Omega,\Omega)\big)-{1 \over 16}\,\epsilon^{ABCD}N^{\bar ba}\pd_a\phi_{AB}\pd_{\bar b}\phi_{CD} \nonumber\\ &&\hspace{0.2 in}+{1 \over 2}\,\psi_A\Box_\Omega\chi^A-{1 \over 4}\,\epsilon^{ABCD}\phi_{AB}\,\Delta(\psi_C,\psi_D)~.
\eea
The fields $\rho$ and $\chi^A$ enter linearly, as Lagrange multipliers for the equations of $\Omega$ and $\psi_A$, while the scalars are paired with themselves. In general the action pairs the component of degree $r$ in $\theta$ with that of degree $4-r$, that is helicity $h$ with helicity $2-h$. It is an action for the positive-helicity multiplet alone, which is possible because Lorentz invariance is already broken by the complex structure.

The equation (\ref{eq:superpleb}) is manifestly invariant under translations of the odd coordinates, $\delta\Phi=\epsilon^A\pd\Phi/\pd\theta^A$. It is also invariant under $\delta\Phi=\epsilon_A^\mu\,\theta^A\pd_\mu\Phi$, since the determinant involves only derivatives with respect to the even coordinates, and the action changes by a total derivative. These transformations generate four scalar and sixteen vector odd charges, $Q_A$ and $S^A_\mu$, with $[Q_A,S^B_\mu]=\delta_A{}^B\pd_\mu$. On one-particle states they act as the supercharges of Section \ref{sec:susy}, up to factors depending on the direction of the momentum.

\subsection{Amplitudes of the superfield theory}
\label{sec:superamplitudes}

We now give some evidence that the superfield theory described above is the correct one. Expanding around the flat background, $\Phi=\eta_{a\bar b}z^a\bz^{\bar b}+\phi$, the kinetic operator is the flat wave operator and the only interaction is cubic. In momentum space, for three one-particle superfields $f_i(\theta)\,e^{ik_i\cdot x}$, the vertex is
\bea
\label{eq:cubicvertex}
V_3={1 \over 2}\,X(1,2)\,\bar X(1,2)\int d^4\theta\,f_1f_2f_3~,
\eea
with $X$ and $\bar X$ as in (\ref{eq:xdef}); by momentum conservation these are the same for each pair of legs. On shell, $X\bar X=(k_1Ik_2)^2/16$ by (\ref{eq:xxbar}), so that the vertex is $(k_1Ik_2)^2$ times a function of the odd variables. To compare with the string amplitude (\ref{eq:superamp}), we first bring the latter to the same form. On the branch $\lambda_i=t_i\lambda$, momentum conservation reads $\sum_it_i\tilde\lambda_i=0$, which implies $[23]/t_1=[31]/t_2=[12]/t_3$. Using these relations and (\ref{eq:rightfactorbranch}) in (\ref{eq:superamp}), one finds
\bea
\label{eq:superampbranch}
A_3=\big(k_1Ik_2\big)^2\;{\delta^{(4)}\big(t_1\eta_1+t_2\eta_2+t_3\eta_3\big) \over Q(\lambda)\,(t_1t_2t_3)^2}~.
\eea
The string amplitude thus contains two factors of $k_1Ik_2$. One is the right-moving factor, and the other is hidden in the left-moving superamplitude. The remaining factor is reproduced by the integral over $\theta$ in (\ref{eq:cubicvertex}) when the one-particle superfields are taken to be $f_i=\sigma_i^{-2}\exp(\sigma_i\,\eta_i\!\cdot\!\theta)$, with $\sigma_i^2=Q(\lambda_i)=t_i^2\,Q(\lambda)$, where $Q$ is the quadratic form of (\ref{eq:rightfactorbranch}). Indeed, in that case $\int d^4\theta\,f_1f_2f_3=(\sigma_1\sigma_2\sigma_3)^{-2}\,\delta^{(4)}(\sigma_1\eta_1+\sigma_2\eta_2+\sigma_3\eta_3)$, which is the second factor of (\ref{eq:superampbranch}). The vertex (\ref{eq:cubicvertex}) and the amplitude (\ref{eq:superamp}) therefore agree, up to a constant. After constant normalizations of the component wavefunctions, the agreement holds for $(\Omega\Omega\rho)$, $(\Omega\phi\phi)$, $(\Omega\psi\chi)$, and $(\phi\psi\psi)$.\footnote{The relative bosonic and fermionic normalizations are fixed by the two-point pairings and supersymmetry Ward identities.} For the bosons, the factors $\sigma_i^{\mp2}$ are precisely the linearized form of the duality of Section \ref{sec:dual}.

Having matched the three-point amplitudes, we turn to higher points. The vertex (\ref{eq:cubicvertex}) depends on the odd variables only through the product $f_1f_2f_3$, and the propagator is $\delta^{(4)}(\theta-\theta')/k^2$. A tree diagram with $n$ legs is then $\int d^4\theta\prod_if_i$ times a function of the momenta alone, and it suffices to study the latter. With the help of AI, we find that the sum over diagrams vanishes on shell for four legs (three diagrams) and for five legs (fifteen diagrams), with no need for contact terms. All four- and five-point tree amplitudes of the sixteen fields therefore vanish, as expected of an $N=2$ string and as anticipated in \cite{Kutasov:1996vh} for the bosons.\footnote{The off-shell form of the vertex matters here. The vertex $(k_1Ik_2)^2/16$, which coincides with $X\bar X$ at three points, gives a nonzero remainder proportional to $stu$ at four points, which must then be cancelled by a quartic term. This is the origin of the contact terms of the non-polynomial action.} The argument of Section \ref{sec:algebra} shows that the same is in fact true for \emph{any} number of legs. As for loops, every closed loop contains a factor $\delta^{(4)}(0)=0$ from the integral over the odd coordinates, reflecting the cancellation $1-4+6-4+1=0$ between the components of the superfield. This is the argument of Siegel for the vanishing of loops in self-dual theories with maximal supersymmetry \cite{Siegel:1992xp,Siegel:1992wd}.\footnote{Note that this does not establish vanishing loop amplitudes of the full string, since charged states can circulate even when every external state is neutral. A direct neutral string four-point calculation with fermions would provide a further test of the proposal, though we do not perform this calculation here.}

\subsection{Relation to earlier work: part I}
\label{sec:knowntheories}

Superfield formulations of self-dual supergravity have a long history, and we close this section by briefly placing (\ref{eq:superpleb}) in context. Siegel showed that the $N=2$ open string is naturally described by self-dual $N=4$ Yang-Mills theory, and gave light-cone superfield actions for self-dual Yang-Mills and supergravity \cite{Siegel:1992xp,Siegel:1992wd}. His action for self-dual $N=8$ supergravity has eight odd coordinates, and he proposed that the $N=(2,1)$ string describes its gauged version.  A covariant form of the $N=8$ action in harmonic superspace was given in \cite{Karnas:1997it}, while twistor actions for self-dual supergravities with any number of supersymmetries were constructed in \cite{Mason:2007ct}. The theory (\ref{eq:superpleb}) differs from these in two respects. First, it has $N=4$ rather than $N=8$ supersymmetry, since the right-movers of the $(5,5)$ string contribute a single state, and do not double the multiplet. Second, it is an action for a single multiplet, with no additional multiplier multiplet, and its field is a K\"ahler potential, in complex coordinates adapted to the complex structure of the $N=2$ string. The second point deserves a comment. In a covariant action of the type of Chalmers and Siegel \cite{Chalmers:1996rq}, the multiplier has the helicity opposite to that of the field whose equation it imposes. Here the multiplier of $\Omega$ is $\rho$, which has helicity $0$ and not $-2$, and belongs to the same multiplet. A covariant description would require a second multiplet, with helicities from $0$ to $-2$. For the open $N=2$ string, a field of this kind was introduced through an additional index on the string field in \cite{Berkovits:1997pq}, and the two fields were related to states in different pictures in \cite{Devchand:1997dq}. Whether the $(5,5)$ string admits a completion of this kind is an open question.

\section{Charged states and the algebra of the left-movers}
\label{sec:charged}

We now turn to the states carrying charge under the lattice $\II_{3,3}$. These include an infinite tower of states which are massless in the base and have unbounded left-moving helicity. We seek a common description of their interactions by a cubic self-dual theory whose couplings are governed by a Lie superalgebra $\g$. Since this algebra is the central object of the section, we begin in Section \ref{sec:idea} with an informal account of what it is, and of why it should be expected to appear. After introducing projected momenta in Section \ref{sec:frame}, we construct an explicit bracket for a maximal isotropic sublattice in Section \ref{sec:tower}. We then test the exchange of higher levels in Section \ref{sec:example}, and in Section \ref{sec:algebra} we explain the underlying BRST algebra and formulate the proposed target theory.

\subsection{The idea: left-moving states as generators}
\label{sec:idea}

Let us first recall how gauge symmetry arises in an ordinary heterotic string. The left-movers contain currents $J^a(z)$, operators of weight one whose zero modes $\oint J^a$ generate a Lie algebra, with structure constants $f^{abc}$ read off from the operator product of two currents. Multiplying a current by a right-moving vector gives the vertex operator of a gauge boson. The gauge bosons are thus in one-to-one correspondence with the generators, and their three-point amplitude is $f^{abc}$ times a kinematic factor supplied by the right-movers. The same is true of the $N=(2,0)$ string, whose target-space theory is self-dual Yang-Mills theory with the gauge algebra of its left-moving currents \cite{Ooguri:1991ie}.

The $(5,5)$ string realizes this mechanism in an extreme form. Its right-movers contribute the same single state to every state of the spectrum, together with the kinematic factor $k_1Ik_2$. Its left-movers are those of a complete $N=1$ string, each of whose physical states $u$ has an integrated vertex operator $V_u(z)$ of weight one. \emph{Every} physical state of the left-movers is therefore in the position of a current, and we may try to define the bracket of two states by the commutator of the corresponding charges,
\bea
\label{eq:chargebracket}
\Big[\oint V_u,\oint V_v\Big]=\oint V_{[u,v]}~.
\eea
This is the construction by which the vertex operators of a lattice generate a Kac-Moody algebra and, when all physical states are included, the Lie algebras of physical states introduced by Borcherds \cite{Borcherds:1983sq}. For the $N=1$ string on the lattice $\II_{9,1}$ it gives the fake monster superalgebra \cite{Scheithauer:1999rp}, which we will meet again in Section \ref{sec:worldline}.

There is one condition for (\ref{eq:chargebracket}) to make sense. The operator product of $V_u(z)$ and $V_v(w)$ contains the factor $(z-w)^{K_u\cdot K_v}$, and the contour integral is defined only if this exponent is an integer. For the internal momenta this is guaranteed by the lattice. For the base momenta it is not, unless they are restricted so that $k_u\cdot k_v=0$ for all pairs, that is to a totally null plane. Remarkably, this is exactly the restriction which the right-movers impose on a three-point amplitude---the three momenta share a spinor, and hence lie in a common null plane, and we will see in Section \ref{sec:frame} that this plane may always be taken to be the holomorphic plane $\Lambda$ of Section \ref{sec:kinematics}. We are thus led to the following definition,
\begin{quote}
\emph{$\g$ is the space of physical states of the left-moving $N=1$ string on $\R^{2,2}\times\II_{3,3}$ whose base momenta lie in $\Lambda$, with the bracket (\ref{eq:chargebracket}).}
\end{quote}
A somewhat more precise version, in terms of BRST cohomology, is given in Section \ref{sec:algebra}. Note the difference from an ordinary heterotic string, whose currents carry no spacetime momentum. Here, because $\Lambda$ is null, the generators depend on two of the four components of the spacetime momentum, and this is how transformations of spacetime itself become part of the gauge algebra.

It is useful to have a picture of what $\g$ looks like. A basis is labelled by the holomorphic components $k_a$ of the base momentum, by the lattice charge $p$, and by a polarization, which distinguishes the different physical states with the same $k_a$ and $p$. The bracket adds momenta and charges,
\bea
\label{eq:grading}
\g=\bigoplus_{k\in\Lambda,\ p\in\II_{3,3}}\g_{(k,p)}~, \hspace{0.4 in} \big[\g_{(k,p)},\g_{(k',p')}\big]\subset\g_{(k+k',p+p')}~.
\eea
Thus $\g$ is infinite-dimensional, and one of its labels is continuous.\footnote{Algebras with continuous labels are familiar, with examples including the Poisson algebra (\ref{eq:poisson}) of functions on a space, and the algebra of gauge transformations of any gauge theory. As in those examples, a complete definition would specify a class of functions of the continuous variables; we do not attempt to do so here.} The polarizations available are those found in Section \ref{sec:spectrum}. In particular, $\g_{(k,p)}$ consists of eight bosons and eight fermions if $p^2=0$, of 128 bosons and 128 fermions if $p^2=-2$, and in general of the states of level $n=-p^2/2$ of the open superstring, and it is empty if $p^2>0$. (These numbers refer to $(k;p)\neq0$; the generators of zero momentum are described in Section \ref{sec:algebra}). In the language of Kac-Moody algebras the pairs $(k;p)$ are the roots and these numbers are their multiplicities. All roots have non-positive norm, since the only state of positive norm would be the Neveu-Schwarz ground state, which is removed by the GSO projection of Section \ref{sec:susy}. The multiplicities are those of the fake monster superalgebra, of which $\g$ is the analogue with the lattice $\II_{9,1}$ replaced by $\Lambda\oplus\II_{3,3}$.\footnote{We do not claim that $\g$ is a generalized Kac-Moody superalgebra in the technical sense: its grading is partly continuous, and its invariant form is indefinite.}

\begin{table}[t]
\centering
\begin{tabular}{@{}llp{6.6cm}l@{}}
\toprule
Charge & States & Bracket & Section\\
\midrule
$p=0$ & $8+8$ & Poisson bracket of functions on $\Lambda$, with values in the functions of four odd coordinates & \ref{sec:neutral}\\\addlinespace
$p$ in an isotropic sublattice & $8+8$ & the same bracket, deformed by the charges as in (\ref{eq:lbracket}) & \ref{sec:tower}\\\addlinespace
$p^2=0$, general & $8+8$ & two charges with $p\cdot p'=-m$ give a state of level $m$ & \ref{sec:example}\\\addlinespace
$p^2=-2n<0$ & $128+128$, \dots & fixed by factorization & \ref{sec:algebra}\\
\bottomrule
\end{tabular}
\caption{The parts of the Lie superalgebra $\g$. The second column gives the number of bosonic and fermionic states for each momentum in $\Lambda$ and each charge.}
\label{tab:galgebra}
\end{table}

The simplest part of $\g$ is the neutral one. A neutral element is a function $f(z,\theta)$ of the two holomorphic coordinates $z^a$ conjugate to $k_a$ and of four odd coordinates, which account for the sixteen polarizations, and its bracket will turn out to be the Poisson bracket $\{f,g\}=\epsilon^{ab}\,\pd_af\,\pd_bg$. The even elements generate the area-preserving diffeomorphisms of the plane. This is the algebra which, used as a gauge algebra, turns self-dual Yang-Mills theory into self-dual gravity; see \cite{Monteiro:2011pc} and references therein. It is the reason why gravity is part of $\g$. Charged elements of level zero behave in the same way as long as their charges are mutually orthogonal, with the bracket deformed by the charges. A new phenomenon occurs for two null charges with $p\cdot p'=-1$. Their sum has norm $-2$, and the bracket of two elements of level zero is then an element of level one, much as the bracket of two root vectors $e_\alpha$ and $e_\beta$ of a simply-laced Lie algebra with $\alpha\cdot\beta=-1$ is the root vector $e_{\alpha+\beta}$. If instead $p\cdot p'=+1$, there are no states of charge $p+p'$ and the bracket vanishes. In this way the null charges generate the higher levels: the excited states of the string are forced upon us by the closure of the algebra. These cases are summarized in Table \ref{tab:galgebra}.

Finally, let us indicate how $\g$ determines a field theory. In a gauge theory the field is a function on spacetime with values in the gauge algebra. Here the algebra already contains the dependence on the holomorphic coordinates, so that a function of the antiholomorphic coordinates $\bz^{\bar a}$ with values in $\g$ is a collection of fields on all of $\R^{2,2}$, one for each charge and polarization. For the neutral part this collection is the superfield $\Phi(z,\bz,\theta)$ of Section \ref{sec:superfield}. The proposal to which this section leads, equation (\ref{eq:TOE}), is that the full target-space theory of the $(5,5)$ string is the self-dual gauge theory of such a field. The remainder of the section assembles the evidence, starting with the left-moving couplings and their dependence on the momenta.

\subsection{The light-cone frame and projected momenta}
\label{sec:frame}

The cubic vertex (\ref{eq:cubicvertex}) of the neutral sector has a suggestive form. Namely, it is the product of a factor $\bar X$ built from the antiholomorphic components of the momenta, which on shell is the three-point factor of the right-movers, and a factor $X\int d^4\theta\,f_1f_2f_3$ built from the holomorphic components, which should then be attributed to the left-movers. To extend this to charged states we need a description of the left-moving couplings in which only the holomorphic components of the base momenta appear. Such a description is provided by light-cone gauge.

Recall from Section \ref{sec:kinematics} that the holomorphic components of a momentum $k$ are its components along the totally null plane $\Lambda$. Let $n$ and $m$ be a basis of the conjugate plane $\bar\Lambda$, and write\footnote{The notation is that of a light-cone frame, in which $k^2=4\,(k_+k_--k_\perp\bar k_\perp)$. The components $k_-$ and $\bar k_\perp$ which are conjugate to $k_+$ and $k_\perp$ in this expression are the antiholomorphic ones, and will not appear in the left-moving couplings.}
\bea
\label{eq:lccomponents}
k_+=n\!\cdot\!k~, \hspace{0.4 in} k_\perp=m\!\cdot\!k~, \hspace{0.4 in} X(1,2)=k_{1+}k_{2\perp}-k_{1\perp}k_{2+}~,
\eea
for the two holomorphic components of $k$ and for the invariant (\ref{eq:xdef}). We use the null vector $n$ to fix a light-cone gauge for the left-movers. A massless vector state of ten-momentum $K=(k;p)$ then has eight polarizations, labelled by vectors $T$ orthogonal to $n$ and defined up to multiples of $n$,
\bea
\label{eq:lcpolarization}
\varepsilon_T(K)=T-{T\!\cdot\!K \over n\!\cdot\!K}\,n~,
\eea
and similarly a massless spinor state is $u=\slashed{K}y$, with $y$ one of eight spinor labels. We work on a patch with $n\cdot K\neq0$. Since $n$ is complex, these polarization vectors are complex even for real momenta. This is only a choice of basis for the space of physical polarizations; a real polarization is a combination of them with complex coefficients, up to a gauge transformation.

The frame has the following property, which is the key to what follows. Consider three on-shell states, whose base momenta share a spinor $\lambda$. On the chosen patch a null rotation in the complexified Lorentz group fixes $n$ and sends the common spinor direction $\lambda$ to the direction $\mu$ defining $\Lambda$. It maps each $k_i$ to the vector $k_i^\Lambda\in\Lambda$ with the same components (\ref{eq:lccomponents}), and transports the polarization at $K_i$ to the corresponding light-cone polarization at $K_i^\Lambda$. The transverse labels are unchanged, up to the gauge freedom along $n$. Spinors and higher oscillator polarizations must be transported in their respective Lorentz representations. Since the left-moving couplings are Lorentz invariant, they are unchanged if each momentum is replaced by
\bea
\label{eq:projected}
K^\Lambda=\big(k^\Lambda;p\big)~.
\eea
In other words, in this frame the left-moving three-point couplings depend on the base momenta only through their holomorphic components $k_+$ and $k_\perp$. Projected momenta have a remarkable feature: since $\Lambda$ is totally null, they are null and mutually orthogonal in the base for \emph{any} set of legs, so that
\bea
\label{eq:integerproducts}
K_i^\Lambda\!\cdot\!K_j^\Lambda=p_i\!\cdot\!p_j\in\Z~.
\eea
The right-moving factor, on the other hand, is not Lorentz invariant, and it is not rotated. It is evaluated at the original momenta, where by (\ref{eq:xxbar}) it is equal to $\bar X(1,2)$ up to a constant and a phase; at projected momenta it would vanish, since $\Lambda$ is a null eigenspace of $I$. We conclude that, up to this constant and phase, the three-point amplitudes of the $(5,5)$ string can be written as
\bea
\label{eq:lcamplitude}
A_3=\bar X(1,2)\ C_L\big(1^\Lambda,2^\Lambda,3^\Lambda\big)~,
\eea
with $C_L$ the left-moving coupling evaluated on projected momenta. For the neutral states the phase is removed by the rescaling of Section \ref{sec:kinematics}. For the charged states the situation is different. On an isotropic sublattice a rescaling which preserves the two-point function exists, and is given in Section \ref{sec:tower}; it acts on the neutral scalars differently from that of Section \ref{sec:kinematics}. For general charges there is none, as two null charges with $p\cdot p'\neq0$ already rule it out.\footnote{In the notation of Section \ref{sec:algebra}, suppose that $U$ is such a rescaling, so that $(Uu,Uv)=(u,v)$ and $([Uu,Uv],Uw)=c\,([u,v],w)$ with $c=e^{-i\vartheta(\lambda)}$. Then $\tilde U=c^{-1}U$ obeys $[\tilde Uu,\tilde Uv]=\tilde U[u,v]$ and $(\tilde Uu,\tilde Uv)=c^{-2}(u,v)$. Applying $\tilde U$ to the relation $[[u_{K,e},u_{-K,e'}],w]=(e\cdot e')(p\cdot p')\,w$, in which $u_{K,e}$ and $w$ have charges $p$ and $p'$, and using the Jacobi identity on its left-hand side, gives $c^2=1$ unless $p\cdot p'=0$.} Let $c=e^{-i\vartheta(\lambda)}$. Rescaling every state by $c$ multiplies the two-point function of the string by $c^2$ and its three-point amplitudes by $c^3$. By (\ref{eq:xxbar}), the latter already carry a factor of $c^{-1}$ relative to those of the theory with $\bar X$, up to a constant. After the rescaling, the quadratic and cubic forms of the string therefore both carry the factor $c^2$ relative to those of that theory, which cancels in the equations of motion. This is the relation between the formulations of self-dual Yang-Mills theory due to Yang \cite{Yang:1977zf} and to Leznov \cite{Leznov:1986up}, whose potentials are related by $J^{-1}\pd_aJ\propto\eta_{a\bar b}\,\epsilon^{\bar b\bar c}\,\pd_{\bar c}\Phi$ and whose cubic vertices contain $k_1Ik_2$ and $\bar X$, respectively. The string amplitudes are of the first kind, and the cubic theory studied below is of the second. For the open $N=2$ string the two formulations have been related to the instanton sectors of the worldsheet gauge field, with the sector of vanishing instanton number giving Yang's, and the sum over all sectors giving Leznov's \cite{Lechtenfeld:1997za}. The corresponding analysis for the $(5,5)$ string is left for the future. For the neutral sector, comparison with (\ref{eq:cubicvertex}) shows that $C_L$ is $X\int d^4\theta\,f_1f_2f_3$, with a suitable dictionary between light-cone states and functions of $\theta$ which we give below.

The form (\ref{eq:lcamplitude}) defines a cubic vertex for arbitrary momenta, on or off shell, and thus a cubic field theory, with propagator $1/k^2$. It is this theory which we will study. That the continuation off shell should be done in this way, rather than for instance by using $k_1Ik_2$ for the right-movers, is not obvious a priori; it will be justified by the results.

\subsection{A maximal isotropic sublattice}
\label{sec:tower}

The simplest charged states are those whose charges lie in a maximal isotropic sublattice $L$ of $\II_{3,3}$, for instance the one spanned by the vectors $\ell_j$ of (\ref{eq:latticebasis}). We write such charges as $p=q_j\,\ell_j/\sqrt2$, with $q_j\in\sqrt2\,\Z$ in the lattice normalization (\ref{eq:latticebasis}). Since all charges in $L$ are null and mutually orthogonal, these states are all at level zero, with sixteen polarizations each, and $L$ is closed under addition, so that they form a closed sector at tree level. Their ten-momenta satisfy $K_i\cdot K_j=k_i\cdot k_j$. We may therefore think of them as the modes of the fields of the neutral sector on a space with three additional coordinates $y_j$, conjugate to the $q_j$, which do not enter the wave operator.

Before turning to the superfield, let us note a simple way of understanding this sector. The rotations of $\mathrm{SO}(5,5)$ generated by $J_{\mu\ell_j}$, which mix a direction $\mu$ of the base with a direction $\ell_j$ of $L$, act on neutral momenta as
\bea
\label{eq:tilt}
(k;0)\ \to\ \big(k;-\omega(k)\big)~, \hspace{0.4 in} \omega\in\mathrm{Hom}\big(\R^{2,2},L\big)~,
\eea
leaving the base momentum, and hence the right-moving factor, unchanged. In position space this is the transformation $x^\mu\to x^\mu+\omega^{\mu j}y_j$. Since the left-moving couplings are invariant under $\mathrm{SO}(5,5)$, a charged amplitude is equal to a neutral amplitude with ``tilted'' polarizations, whenever the charges $q_i$ of all legs can be written as $-\omega(k_i)$ for a single $\omega$. For $n$ legs there are $n-1$ independent momenta, and such an $\omega$ exists as long as these are linearly independent in the four-dimensional base. Thus any configuration of four or five legs with charges in $L$ is a tilt of a neutral one, while for six or more legs this is no longer the case. An immediate consequence is that the four- and five-point amplitudes of the string in this sector vanish, to all orders in $\alpha'$, if the neutral ones do.

\subsubsection{Charged vertices and a Lie superalgebra} 
We now give the left-moving coupling explicitly. Let us label the four odd coordinates of Section \ref{sec:superfield} as $(\theta^0,\theta^j)$, $j=1,2,3$, in a basis adapted to $L$, and write $\pd_0,\pd_j$ for the derivatives with respect to them. As in Section \ref{sec:superamplitudes}, the sixteen states of a leg are described by a function of the odd coordinates. We denote it by $F(\theta)$, with a capital letter to distinguish the normalization used here from that of Section \ref{sec:superamplitudes}, and write $F_i$ for the function of the $i$-th leg. The light-cone states of Section \ref{sec:frame} are mapped to functions of $\theta$ as follows. First, the two polarizations along the base are mapped to $1/k_+$ and to ${1 \over 2}k_+\,\theta^0\theta^1\theta^2\theta^3$. Second, the scalars along $\ell_j$ and along $\ell'_j$ are mapped to $\epsilon_{jkl}\theta^k\theta^l/\sqrt2$ and to $\theta^0\theta^j/\sqrt2$. Finally, the spinors are mapped to the linear and cubic monomials. The sublattice $L$ thus singles out one of the four odd coordinates: $L$ corresponds to the bivectors not involving $\theta^0$, and the dual sublattice $L'$ to those involving it. With this dictionary one finds, for neutral legs, that all of the left-moving couplings of three vectors and of two spinors and a vector are reproduced by $X\int d^4\theta\,F_1F_2F_3$, as anticipated. For legs with charges in $L$, on the other hand, we find that the left-moving coupling is
\bea
\label{eq:chargedvertex}
C_L=\int d^4\theta\,[F_1,F_2]_L\,F_3~,
\eea
where the bracket of two superfields with holomorphic momenta $(k_{i+},k_{i\perp})$ and charges $q_i$ is
\bea
\label{eq:lbracket}
[F_1,F_2]_L&=&X\,F_1F_2-{(D_{12}F_1)\,F_2 \over k_{1+}}-{F_1\,(D_{12}F_2) \over k_{2+}}+{D_{12}(F_1F_2) \over k_{1+}+k_{2+}}~,
 \nonumber\\ D_{12}&=&\big(k_{1+}q_{2j}-k_{2+}q_{1j}\big)\,\pd_0\pd_j~.
\eea
This formula reproduces every component of the bosonic and fermionic cubic couplings that we have checked. The tilt (\ref{eq:tilt}) explains why the charges appear only to first order. The coupling (\ref{eq:chargedvertex}) is the neutral coupling $X\int d^4\theta\,F_1F_2F_3$ with each $F_i$ replaced by the tilted state $\big(1-D_{12}/(Xk_{i+})\big)F_i$, where the operator $D_{12}$ squares to zero. In the product of the three tilted states, the term of third order in the charges vanishes because it does not contain $\theta^0$, and the terms of second order cancel between the three pairs of legs because $k_{1+}+k_{2+}+k_{3+}=0$. The charges thus enter through a single second-order differential operator in the odd coordinates. 

Note that for Grassmann-even superfields, including odd component fields, the bracket (\ref{eq:lbracket}) is antisymmetric, its cyclic trilinear form is totally antisymmetric, and it satisfies the Jacobi identity
\bea
\label{eq:jacobi}
\big[[F_1,F_2]_L,F_3\big]_L+\big[[F_2,F_3]_L,F_1\big]_L+\big[[F_3,F_1]_L,F_2\big]_L=0~.
\eea
It therefore defines a Lie superalgebra on the space of functions of the two holomorphic coordinates, of the three additional coordinates $y_j$, and of the four odd coordinates, with an invariant inner product. For vanishing charges it is the Poisson algebra of the plane $\Lambda$ with values in the functions of $\theta$. Note that the charges do not simply multiply $X$ by the structure constants of some internal algebra, as one might have guessed: they deform the Poisson bracket itself. The identity (\ref{eq:jacobi}) is quite non-trivial, and fails if the relative coefficients of the three terms involving $D_{12}$ in (\ref{eq:lbracket}) are changed.

 Combining (\ref{eq:chargedvertex}) with the right-moving factor as in (\ref{eq:lcamplitude}), we obtain a cubic theory for superfields depending on $(x^\mu,y_j,\theta^A)$.\footnote{On the tower the phase in (\ref{eq:xxbar}) can be absorbed without changing the two-point function, by the rescaling $T_cF(\theta)=c\,F(c\,\theta^0,c^{-1}\theta^1,c^{-1}\theta^2,c^{-1}\theta^3)$ with $c=e^{-i\vartheta(\lambda)}$. Indeed, the operators $\pd_0\pd_j$ commute with $T_c$, one has $(T_cF)(T_cG)=c\,T_c(FG)$, and the rescaling of the odd coordinates has determinant $c^{-2}$, so that $\int d^4\theta\,(T_cF)(T_cG)=\int d^4\theta\,FG$ while the cubic form (\ref{eq:chargedvertex}) is multiplied by $c$. On neutral legs $T_c$ differs from the rescaling of Section \ref{sec:kinematics} by a complexified rotation of the internal space, which multiplies the scalars along $L'$ by $c$ and those along $L$ by $c^{-1}$, and under which the neutral couplings are invariant.} In position space, the cubic theory is described by a direct generalization of (\ref{eq:superpleb}),
\bea
\label{eq:toweraction}
S_L={1 \over 2}\int d^4x\,d^3y\,d^4\theta\,\Big[\Phi+{1 \over 3}\,\Phi\,{\det}_L\Phi\Big]~, \hspace{0.4 in} {\det}_L\Phi={1 \over 2}\,\epsilon^{\bar a\bar b}\,\big[\pd_{\bar a}\Phi,\pd_{\bar b}\Phi\big]_L~,
\eea
which reduces to (\ref{eq:superpleb}) for fields independent of $y$. The bracket of the coordinate functions $z^a$ with any field is not deformed, the terms involving the charges in (\ref{eq:lbracket}) cancelling against each other, so that the expansion of $\det_L\Phi$ around flat space has the ordinary wave operator of the base as its kinetic term, as it should.

As described in the acknowledgements, using AI we have computed the sum of all cubic tree diagrams of this theory with four, five, six, and seven external legs, for generic charges in $L$ and generic external superfields, and find that it vanishes in each case. Recall that for six or more legs the configurations are not tilts of neutral ones, so that this result is not a consequence of the vanishing of the neutral amplitudes. An argument for any number of legs is given in Section \ref{sec:algebra}. If on the other hand the right-moving factor is taken to be $k_1Ik_2$, continued off shell in the obvious way, the trees with charged legs do not vanish, although those with neutral legs do. This is the justification promised above for the form (\ref{eq:lcamplitude}).

\subsection{Timelike charges: an example}
\label{sec:example}

The isotropic sublattice is special in that all of its states are at level zero. We now ask what happens when the charges are such that states of higher level appear. The simplest example is provided by four states with the null charges
\bea
\label{eq:examplecharges}
p_1=\ell_1~, \hspace{0.4 in} p_2=\ell'_1~, \hspace{0.4 in} p_3=-\ell_1~, \hspace{0.4 in} p_4=-\ell'_1~.
\eea
Each of these states is at level zero. But $p_4+p_1=\ell_1-\ell'_1$ has norm $-2$, so that the fourth and first states can fuse into a state of level one, with 256 polarizations; $p_4+p_2=0$, so that the fourth and second states fuse into a neutral state; and $p_4+p_3$ has norm $+2$, for which there are no states at all. A four-point tree amplitude with these legs thus receives contributions from the exchange of the level-one multiplet in one channel and of the neutral multiplet in another, and it is far from obvious that the two should cancel.

\begin{figure}[t]
\centering
\begin{tikzpicture}[font=\small]
\draw[gray!60] (-4.2,-1.7) rectangle (4.2,1.9);
\node[gray] at (3.85,1.6) {$z$};
\fill (-2.6,0.2) circle (2pt) node[below=3pt] {$z_1$};
\fill (0,-0.6) circle (2pt) node[below=3pt] {$z_2$};
\draw (2.6,0.4) circle (2pt) node[below=3pt] {$z_3$};
\draw[blue!60!black] (-2.6,0.2) circle (0.55);
\draw[blue!60!black] (0,-0.6) circle (0.55);
\node[align=center] at (-2.6,1.25) {level 1\\[-2pt] ($K_4\!\cdot\!K_1=-1$)};
\node[align=center] at (0,0.45) {level 0\\[-2pt] ($K_4\!\cdot\!K_2=0$)};
\node[align=center] at (2.6,1.1) {no pole\\[-2pt] ($K_4\!\cdot\!K_3=+1$)};
\end{tikzpicture}
\caption{The left-moving four-point function at projected momenta, as a function of the position $z$ of the fourth vertex operator, for the charges (\ref{eq:examplecharges}). It is a rational function with poles at $z_1$ and $z_2$ only, whose residues are the exchanges of the level-one and of the neutral multiplet. The two residues add up to zero.}
\label{fig:poles}
\end{figure}
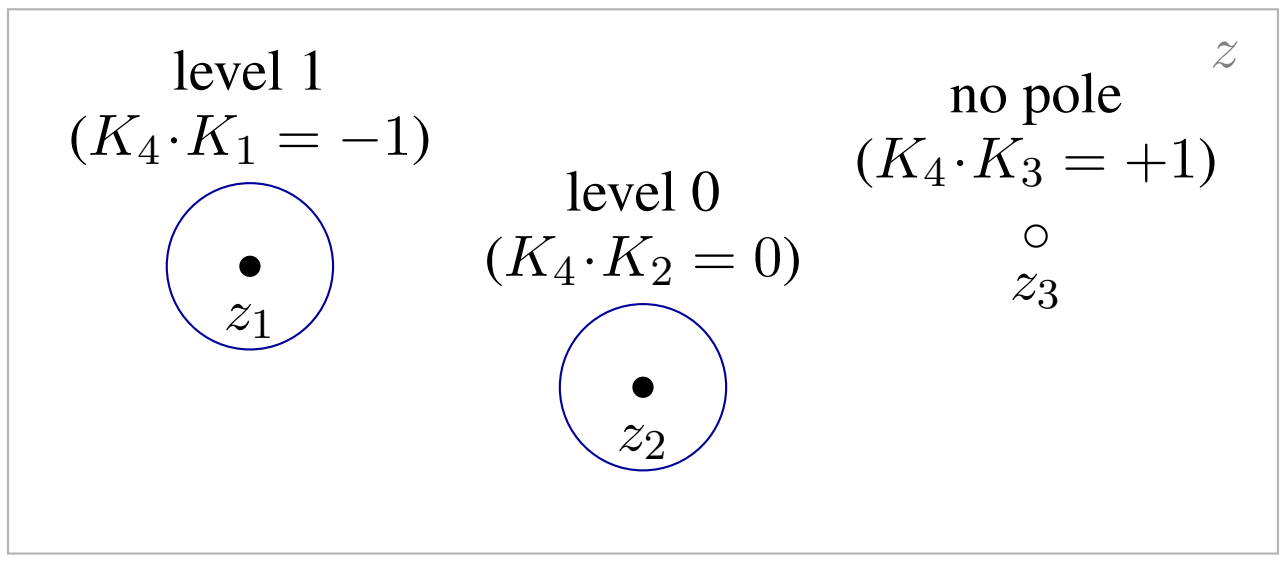

To see that they do, consider the left-moving four-point function of the vertex operators of the four states, evaluated at the projected momenta (\ref{eq:projected}), as a function of the position $z$ of the fourth operator with the other three held fixed. We take all four states to be bosons, with two of the operators in the $-1$ picture and two in the $0$ picture; the details are given in Appendix \ref{app:correlators}. Since the fields are free, the correlator is the product of a factor
\bea
\label{eq:koba}
\prod_{i<j}(z_i-z_j)^{K_i^\Lambda\cdot K_j^\Lambda}
\eea
and a rational function of the positions. By (\ref{eq:integerproducts}), the exponents in (\ref{eq:koba}) are integers, i.e. $K_4\cdot K_1=-1$, $K_4\cdot K_2=0$, and $K_4\cdot K_3=+1$. The four-point function $A_L(z)$ is therefore a rational function of $z$, which can have poles only where the fourth operator meets one of the other three. Near $z_i$ it is governed by the operator product of the fourth and $i$-th operators: a pole with $K_4\cdot K_i=-n$ has as its residue the contribution of the physical states of level $n$ exchanged in that channel. In the present example there is a pole at $z_1$, where the level-one multiplet is exchanged, and a pole at $z_2$, where the neutral multiplet is exchanged. There is no pole at $z_3$, since $K_4\cdot K_3=+1$ and there are no physical states of charge $p_4+p_3$. Finally, since the fourth operator has weight one, $A_L(z)$ falls off as $1/z^2$ as $z\to\infty$, so that $A_L(z)\,dz$ has no pole at infinity. This is illustrated in Figure \ref{fig:poles}.

The sum of the residues of a rational function vanishes. Hence
\bea
\label{eq:residuesum}
\Res_{z\to z_1}A_L(z)+\Res_{z\to z_2}A_L(z)=0~,
\eea
and the exchange of the level-one multiplet between the pairs $(4,1)$ and $(2,3)$ is equal and opposite to the exchange of the neutral multiplet between the pairs $(4,2)$ and $(3,1)$. Denoting by $[u,v]$ the state obtained by fusing two physical states, which for two massless vectors with $K_u\cdot K_v=0$ is the massless vector with polarization $(e_u\!\cdot\!K_v)\,e_v-(K_u\!\cdot\!e_v)\,e_u+(e_u\!\cdot\!e_v)\,K_u$ and for $K_u\cdot K_v=-1$ is a specific state of the first massive level given in Appendix \ref{app:levelone}, we find
\bea
\label{eq:factorization}
\Res_{z\to z_i}A_L(z)=\pm\big([u_4,u_i],[u_j,u_k]\big)~,
\eea
where $(u,v)$ denotes the inner product of two physical states of opposite momenta and charges. For two massless vectors with polarizations $e$ and $e'$ it is $e\cdot e'$, and for two states of level one it is given in (\ref{eq:invariantform}). It is called invariant because $([u,v],w)=(u,[v,w])$, both sides being the three-point coupling of $u$, $v$, and $w$. The signs in (\ref{eq:factorization}) are the ordering signs of the lattice vertex operators, which are given in (\ref{eq:factorizationsigns}). In particular, the residue at the level-one pole is the inner product of two explicit level-one states. As the polarizations of the two fusing bosons are varied, these states span 64 of the 128 bosons of the level.

It remains to include the right-movers. In the cubic theory defined by (\ref{eq:lcamplitude}), the four-point tree amplitude is the sum over the three channels of the product of the left-moving residue, the right-moving factors of the two vertices, and the propagator. The right-moving part of each term is
\bea
\label{eq:rightidentity}
{\bar X(4,i)\,\bar X(j,k) \over s_{4i}}~, \hspace{0.4 in} s_{4i}=2\,k_4\!\cdot\!k_i~,
\eea
with $(i,j,k)$ a cyclic permutation of $(1,2,3)$. For four null momenta adding up to zero, this quantity is the same in all three channels; this is an elementary identity, which in spinor variables states that $[jk]/\langle4i\rangle$ is independent of the channel.\footnote{To see this, note that by (\ref{eq:xspinor}), and since $s_{4i}=\langle4i\rangle[4i]$, the quantity (\ref{eq:rightidentity}) is equal to ${1 \over 64}\big(\prod_{l=1}^4\langle\mu l\rangle\big)\,[jk]/\langle4i\rangle$, whose first factor is symmetric in the four legs. Contracting the equation $\sum_l\lambda_{l\alpha}\tilde\lambda_{l\dot\alpha}=0$ with $\lambda_4$ and $\tilde\lambda_j$ gives $\langle4i\rangle[ij]+\langle4k\rangle[kj]=0$, that is $[jk]/\langle4i\rangle=[ij]/\langle4k\rangle$. The right-hand side is the same ratio for the channel in which the fourth leg fuses with leg $k$, the pair $(i,j)$ again being in cyclic order. 
} The four-point amplitude is therefore a common factor times the sum of residues (\ref{eq:residuesum}), and vanishes. Note that both ingredients are needed---if the level-one states are omitted, the four-point amplitude of the remaining theory is nonzero. Thus the level-zero states with general null charges do not form a consistent truncation. 

\subsection{The Lie superalgebra and the target-space theory}
\label{sec:algebra}

It is clear that the argument just given is not special to the charges (\ref{eq:examplecharges}), nor to the states considered. For any four physical vertex operators of the left-moving string, evaluated at projected momenta, the exponents (\ref{eq:integerproducts}) are integers, the correlator is a rational function of the position of an integrated vertex operator with poles only at the other insertions, and the residues add up to zero. In the Ramond sector the individual factors have square-root branch cuts, but these cancel between the spin fields, the fermions, and the superghosts, as a consequence of the GSO projection. Combined with the identity for (\ref{eq:rightidentity}), which involves only the base momenta, this shows that the four-point tree amplitudes of the cubic theory vanish for states of all levels.

We can now make precise the definition of $\g$ sketched in Section \ref{sec:idea}, and restate the results above in its terms. The construction is the standard one by which the physical states of a chiral string form a Lie algebra \cite{Borcherds:1983sq}, in the form appropriate to the superstring, which is described in \cite{Scheithauer:1999rp}. The elements of $\g$ are the physical states of the left-moving $N=1$ string, with the GSO projection, whose base momenta lie in $\Lambda$. As usual, two physical states are identified if they differ by a null state, that is by a physical state which is orthogonal to all physical states.\footnote{Since the inner product is indefinite, there are also physical states of zero norm which are not of this kind. These are kept.} The bracket of two states is defined by (\ref{eq:chargebracket}), and involves integrating the vertex operator of the first state around that of the second. Here the vertex operator of a state of charge $p$ includes the standard cocycle factor of the lattice \cite{Scheithauer:1999rp}, as a result of which the product of two operators of charges $p$ and $q$ acquires a sign $\epsilon(p,q)$, with $\epsilon(p,q)/\epsilon(q,p)=(-1)^{p\cdot q}$. Without this factor, two operators with $p\cdot q$ odd would anticommute, and their contour product would be symmetric where the bracket must be antisymmetric. By (\ref{eq:integerproducts}) the operator product of the two vertex operators contains only integer powers of $z-w$, so that the integral is well-defined. 

For two Neveu-Schwarz states, represented by operators $u$ and $v$ of weight ${1 \over 2}$ built from the matter fields, the result is
\bea
\label{eq:gbracket}
[u,v]=\big(G_{-1/2}u\big)_0\,v~,
\eea
where $G_{-1/2}u$ is the operator of weight one which appears in the vertex operator of $u$ in the $0$ picture, and the subscript denotes its zero mode. If one or both of the states are in the Ramond sector the definition is the same, with the vertex operators taken in suitable pictures.

The bracket of two physical states is again a physical state, since the integrated vertex operator of a physical state commutes with the BRST charge. Its antisymmetry and its Jacobi identity, with the usual signs when fermions are involved, follow by deforming contours. The vanishing of the sum of residues in Section \ref{sec:example} is the same statement, i.e. by (\ref{eq:factorization}), it is the Jacobi identity for $u_1$, $u_2$, and $u_3$, paired with $u_4$. For the brackets to close, $\g$ must also contain ten even generators $P^M$ of zero momentum, represented by the operators $i\pd X^M$: for two massless vectors of opposite momenta, the bracket of Appendix \ref{app:levelone} gives $[u_{K,e},u_{-K,e'}]=(e\cdot e')\,K_MP^M$. Since $P^M$ multiplies a state of momentum $R$ by $R^M$, the right-hand side acts as $K\cdot R$, which for projected momenta is the product $p\cdot r$ of the two charges and need not vanish. These generators are the analogue of the Cartan subalgebra of a Kac-Moody algebra. For two neutral states the right-hand side acts as zero on all of $\g$, and is central. The description of the neutral part of $\g$ as a Poisson algebra, in which the bracket of two plane waves of opposite momenta vanishes, is to be understood modulo these central elements. Following \cite{Scheithauer:1999rp}, we take the zero-momentum part of $\g$ to consist of the ten even generators $P^M$ alone. The sixteen Ramond states of zero momentum, which are the supercharges of Section \ref{sec:susy}, are not elements of $\g$, but act on it.

The results of this section can now be summarized as follows.
\begin{itemize}
\item The left-moving three-point couplings of the $(5,5)$ string, in the frame of Section \ref{sec:frame} and with the cocycle factors included, are the structure constants of $\g$;
\item The part of $\g$ of vanishing charge is the Poisson algebra of functions on the plane $\Lambda$, with values in the functions of the four odd coordinates $\theta^A$. It is this algebra which, combined with the Poisson algebra of the right-movers, gives the self-dual supergravity of Section \ref{sec:neutral};
\item The part of $\g$ with charges in a maximal isotropic sublattice is the algebra defined by the bracket (\ref{eq:lbracket}). The Jacobi identity (\ref{eq:jacobi}), which there appeared as the result of a computation, is an instance of the general statement;
\item The higher oscillator levels supply the states required for closure when charges outside an isotropic sublattice are combined. 
\end{itemize}

We are thus led to the following proposal: \emph{the tree-level target-space theory of the $(5,5)$ string is the cubic theory of a field $\Phi$ on $\R^{2,2}$ with values in $\g$}, with equation of motion
\bea
\label{eq:leznov}
\Box\,\Phi+{1 \over 2}\,\epsilon^{\bar a\bar b}\,\big[\pd_{\bar a}\Phi,\pd_{\bar b}\Phi\big]_\g=0~,
\eea
up to normalization. Note that $\Phi$ is no longer the superfield of Section \ref{sec:neutral}, but a much larger object, of which that superfield is one component. It is a sum $\Phi=\sum_p\Phi_p$ over the charges $p\in\II_{3,3}$ with $p^2\leq0$, in which $\Phi_p$ is a field on $\R^{2,2}$ with as many components as there are states of charge $p$, i.e. sixteen for null $p$, and 256 or more for timelike $p$. The component $\Phi_0$ is the superfield of Section \ref{sec:neutral}, with the flat background subtracted, and the components with charges in $L$ make up the field of (\ref{eq:toweraction}). The bracket in (\ref{eq:leznov}) maps the components of charges $p$ and $p'$ to that of charge $p+p'$. Here the dependence of $\Phi$ on the holomorphic coordinates is part of the structure of $\g$, while the antiholomorphic derivatives are contracted as in the Poisson bracket of the right-movers. Equation (\ref{eq:leznov}) has the form of self-dual Yang-Mills theory in the cubic formulation of Leznov and Parkes \cite{Leznov:1986up,Parkes:1992rz}, with $\g$ as gauge algebra, and it follows from the action 
\bea
\label{eq:TOE}
S={1 \over 2}(\Phi,\Box\Phi)+{1 \over 6}\epsilon^{\bar a\bar b}(\Phi,[\pd_{\bar a}\Phi,\pd_{\bar b}\Phi]_\g)~,
\eea
 with $(\ ,\ )$ the invariant inner product defined above. For the neutral part of $\g$, it is the expansion of (\ref{eq:superpleb}) around flat space, and for charges in an isotropic sublattice it is that of (\ref{eq:toweraction}).

The evidence for this proposal is that it reproduces the three-point amplitudes of the string by construction (up to the phase discussed in Section \ref{sec:frame}, which for general charges cannot be absorbed in the normalization of the states), and that its four-point tree amplitudes vanish, as shown above. On the isotropic tower the tree cancellations have been checked through seven points using AI. In fact the tree amplitudes vanish beyond three points for any number of legs, at generic momenta, by the argument of Section 4 of \cite{Monteiro:2022lwm}.\footnote{We thank Ricardo Monteiro for bringing this to our attention.} In the present setting it runs as follows. The theory (\ref{eq:TOE}) is self-dual Yang-Mills theory with gauge algebra $\g$, and each of its cubic diagrams is the product of a factor built from the brackets of $\g$ and a factor built from $\bar X$, divided by the propagators. As for an ordinary gauge algebra, the Jacobi identity of $\g$ allows the sum over diagrams to be written as a sum of color-ordered amplitudes of self-dual Yang-Mills theory, with coefficients built from the brackets of $\g$, and these amplitudes vanish beyond three points. The only new feature is that the brackets of $\g$ depend on the momenta. They do so only through the projected momenta, for arbitrary values of which the Jacobi identity holds, so that no restriction to collinear momenta is needed.

The four-point amplitudes of the string itself also vanish at generic momenta, for charged as well as neutral states. For neutral bosonic states this is known \cite{Ooguri:1991ie,Kutasov:1996vh}, and the argument involves only the right-movers. With three of the vertex operators at $0$, $1$, and infinity, the integrand of the sphere amplitude is a sum of terms $z^{a}(1-z)^{b}\,\bar z^{\bar a}(1-\bar z)^{\bar b}$ in which $a-\bar a$ and $b-\bar b$ are integers. The charges change the left-moving exponents only by the integers $p_i\cdot p_j$, so that this remains true for charged states, and each term is single-valued. The integral of such a term over the plane is the product of the left-moving integral $\int_0^1dx\,x^{a}(1-x)^{b}$, the right-moving integral $\int_0^1dx\,x^{\bar a}(1-x)^{\bar b}$, and a factor which is the same for all terms \cite{Kawai:1985xq}. The sum of the right-moving integrals over the terms is a four-point amplitude of the open $N=2$ string, which vanishes for null momenta in $\R^{2,2}$ \cite{Ooguri:1990ww,Ooguri:1991fp}. The amplitude of the $(5,5)$ string therefore vanishes whatever its left-moving factor is. This is the statement in the sector of vanishing instanton number, and it extends to the other sectors by spectral flow \cite{Lechtenfeld:1995zr}.

\subsection{Relation to earlier work: part II}
\label{sec:earlierwork}

That $N=2$ heterotic strings describe self-dual gauge fields whose gauge algebra is determined by the left-movers is of course not new. It was shown in \cite{Ooguri:1991ie} that the $N=(2,0)$ string describes self-dual Yang-Mills theory, reduced to two or three dimensions, with gauge group given by the current algebra of the internal lattice, coupled to gravity; and the appearance of generalized Kac-Moody algebras in the $N=(2,1)$ string, and their possible relation to algebras of BPS states \cite{Harvey:1995fq}, was discussed in \cite{Kutasov:1996vh}. What the $(5,5)$ string adds to this picture is the following. Since no null reduction is imposed, the gauge algebra is not that of the massless states of an internal theory, but the full algebra $\g$ of physical states of the left-movers, including the dependence on two of the four spacetime momenta. As a result, self-dual gravity and self-dual gauge theory are not coupled to each other as separate sectors---the gravity multiplet is the neutral part of the same algebra whose charged parts are the gauge and higher-helicity multiplets. The proposed description is cubic, as in \cite{Parkes:1992rz}. The exact neutral duality illustrates why a non-polynomial action in another choice of potentials can describe the same classical system. It would be interesting to understand if there is any relation to the four-dimensional current algebras of \cite{Losev:1995cr}.

\section{Other vacua}
\label{sec:vacua}

The $(5,5)$ string is one member of the family of theories introduced in Section \ref{sec:realforms}. In this section we briefly describe some of the other members, showing how their left-moving physical-state algebras, null reductions, and right-moving kinematics determine their spectra and cubic couplings.

\subsection{Vacua as isotropic subspaces}
\label{sec:isotropic}

Recall that a member of the family is specified by an even self-dual lattice $\II_{a,b}$ and by $g$ mutually orthogonal null vectors in $\R^{2,2}\times\II_{a,b}$, which span an isotropic subspace. Using the lattice quotient of Section \ref{sec:nullghosts}, internal null gaugings can be traded for a smaller lattice. Gaugings involving the base also constrain the shared momenta. The intersection of the gauged isotropic subspace with the base therefore distinguishes the resulting constructions. For the standard string, with $\sigma=8$, the possibilities are listed in Table \ref{tab:vacua}. With one null current we recover the two vacua of Section \ref{sec:heteroticreview}. With two null currents and the lattice $\II_{9,1}$, a choice of one base null vector and one mixed null vector gives the worldline vacuum described below. Taking both null vectors to span a maximal isotropic plane entirely within the base instead gives a point vacuum, whose possible matrix interpretation we discuss in Section \ref{sec:point}.

\begin{table}[t]
\centering
\begin{tabular}{@{}llll@{}}
\toprule
$\sigma$ & Lattice, null currents & Null vectors & Target space\\
\midrule
$8$ & $E_8$, $g=1$ & one in the base & string, $1+1$\\
$8$ & $E_8$, $g=1$ & one, timelike part in the base & membrane, $2+1$\\
$8$ & $\II_{9,1}$, $g=2$ & one in the base, one mixed & worldline, $0+1$\\
$8$ & $\II_{9,1}$, $g=2$ & two in the base & point\\
$0$ & $\II_{3,3}$, $g=0$ & none & $2+2$\\
\bottomrule
\end{tabular}
\caption{Some vacua of the $N=(2,1)$ string. The target dimensions refer to the reduced sectors. }
\label{tab:vacua}
\end{table}

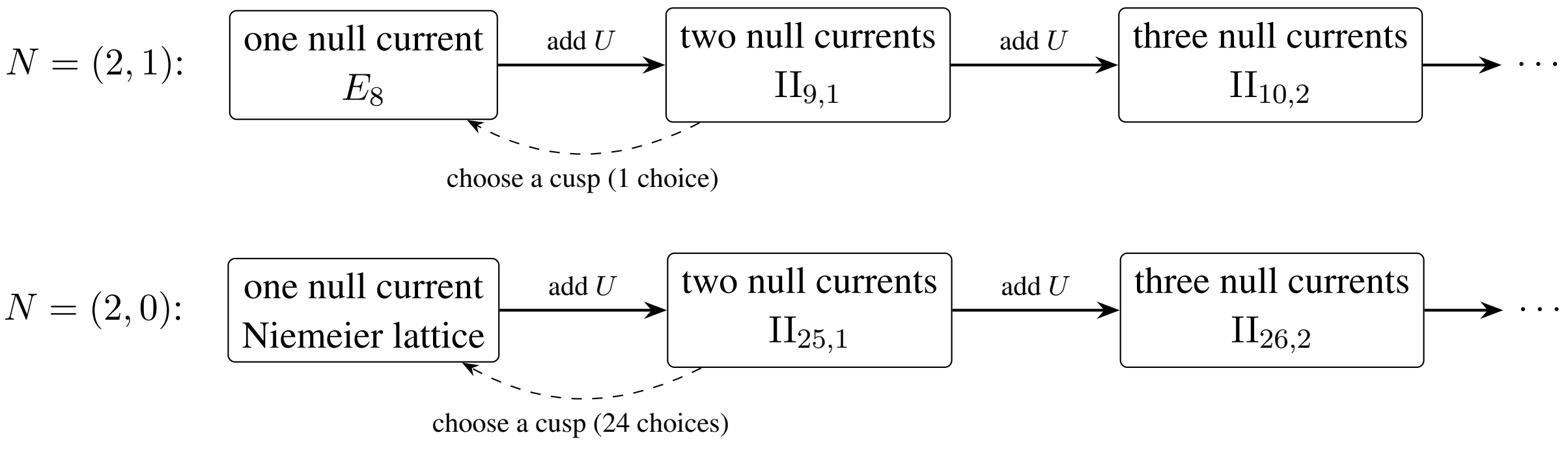
\begin{figure}[t]
\centering
\begin{tikzpicture}[node distance=14mm and 17mm, every node/.style={font=\small}, box/.style={draw,rounded corners=2pt,inner sep=4pt,align=center}]
\node[box] (a1) {one null current\\ $E_8$};
\node[box,right=of a1] (a2) {two null currents\\ $\II_{9,1}$};
\node[box,right=of a2] (a3) {three null currents\\ $\II_{10,2}$};
\node[left=3mm of a1] {$N=(2,1)$:};
\node[box,below=of a1] (b1) {one null current\\ Niemeier lattice};
\node[box,right=of b1] (b2) {two null currents\\ $\II_{25,1}$};
\node[box,right=of b2] (b3) {three null currents\\ $\II_{26,2}$};
\node[left=3mm of b1] {$N=(2,0)$:};
\draw[-{Stealth},thick] (a1) -- node[above,font=\scriptsize]{add $U$} (a2);
\draw[-{Stealth},thick] (a2) -- node[above,font=\scriptsize]{add $U$} (a3);
\draw[-{Stealth},thick] (b1) -- node[above,font=\scriptsize]{add $U$} (b2);
\draw[-{Stealth},thick] (b2) -- node[above,font=\scriptsize]{add $U$} (b3);
\node[right=8mm of a3] (a4) {$\cdots$};
\node[right=8mm of b3] (b4) {$\cdots$};
\draw[-{Stealth},thick] (a3) -- (a4);
\draw[-{Stealth},thick] (b3) -- (b4);
\draw[-{Stealth},dashed] (a2) to[bend left=28] node[below,font=\scriptsize]{choose a cusp (1 choice)} (a1);
\draw[-{Stealth},dashed] (b2) to[bend left=28] node[below,font=\scriptsize]{choose a cusp (24 choices)} (b1);
\end{tikzpicture}
\caption{Adding a hyperbolic plane $U$ and a null current enlarges the lattice. Conversely, a choice of primitive null vector in the larger lattice determines the smaller one.}
\label{fig:ladder}
\end{figure}

One may continue in this way, adding hyperbolic planes to the lattice and null currents to the gauge algebra, as in Figure \ref{fig:ladder}. Going up, the lattice becomes larger and more symmetric; going down, one must choose a primitive null vector $v$ of the lattice, and the theory one lands on is the one with lattice $v^\perp/v$. From this point of view the null vector is a choice of cusp of the larger lattice, that is, of an orbit of primitive null vectors under its automorphism group.

\subsection{The worldline vacuum and the fake monster superalgebra}
\label{sec:worldline}

Consider left-movers on $\R^{2,2}\times\II_{9,1}$, of total signature $(11,3)$, with two null supercurrents. We take the first null vector $v_1$ to lie in the base, and the second to be $v_2=\hat\tau+x_2$, with $x_2$ a spacelike direction of the base and $\hat\tau$ a unit timelike vector in $\II_{9,1}\otimes\R$. In the reduced sector, the base motion is restricted to a worldline, with energy determined by $\hat\tau$. We will also keep track of the additional null label present before the zero-mode identification.

The neutral states are easily found. In the Neveu-Schwarz sector there are $14-2-2=10$ massless bosons, which may be thought of as a gauge field $A_0$ and nine scalars. In the Ramond sector the two null constraints and the Dirac equation reduce the 64 components of a spinor to 16, forming an irreducible spinor of $\mathrm{Spin}(9)$. This is the field content of the worldline theory of a D-particle, with sixteen supercharges. Charged states exist for every nonzero $p\in\II_{9,1}$ with $p^2\leq0$. They have energy $E=|p\cdot\hat\tau|$, and at level $n=-p^2/2$ their numbers are $8,128,1152,\dots$ for bosons and fermions alike.

This is a familiar structure. The lattice $\II_{9,1}$ is the root lattice of $E_{10}$, and the states just described are the roots of the \emph{fake monster superalgebra} of Scheithauer \cite{Scheithauer:1999rp}, the Lie superalgebra of physical states of the chiral $N=1$ string on $\II_{9,1}$, with the multiplicities proven there.\footnote{The name is inherited from the bosonic string. The Lie algebra of physical states of the bosonic string on the lattice $\II_{25,1}$ was originally called the monster Lie algebra, since it was expected to be related to the monster group. When the algebra which carries an action of the monster group, and which enters the proof of the moonshine conjectures, was found, it took over that name, and the earlier algebra became the fake monster Lie algebra. The fake monster superalgebra is the analogue of the latter for the superstring.} That a superalgebra of this kind should be related to the $N=(2,1)$ string was anticipated in \cite{Kutasov:1996vh}.

It can be checked that the left-moving three-point couplings of this vacuum reproduce the bracket of the superalgebra. For two null roots $\alpha$ and $\beta$ with $\alpha\cdot\beta=-1$, the fusion of two bosonic states lands in the first massive level, whose 128 bosons now have positive norm; the 64 states obtained in this way are physical and linearly independent, and their inner products computed in the worldline vacuum, with both null supercurrents imposed, agree with those computed directly in the lattice vertex algebra. The eight neutral supercharges which act on a null root, and the sixteen which act on a root of level one, organize the states into multiplets of $8+8$ and $128+128$.

The right-moving factor depends on the zero-mode prescription. Retain the null label before the additional identification, choose $v_1=t^1+x^1$, and write $c_i=p_i\cdot\hat\tau$. On the branch with base momenta $k_i=(a_i,c_i,a_i,-c_i)$ in the coordinate order of Appendix \ref{app:conventions}, one finds
\bea
\label{eq:worldlineright}
k_1Ik_2=2\,(a_1c_2-a_2c_1)~.
\eea
Here $a_i$ is the retained null label, which is the momentum conjugate to the null coordinate $u=v_1\cdot x=x^1-t^1$, while $c_i$ is conjugate to the worldline coordinate $t^2+x^2$. In the sector in which the labels $a_i$ vanish, which is the one usually considered, every three-point amplitude vanishes, and the superalgebra is visible only in the left-moving couplings. For nonzero labels the amplitudes are nonzero, and the fields of the Cartan subalgebra, namely $A_0$ and the scalars, couple to a root $p$ through their derivative along $u$ with the expected charge. These nonzero cubic couplings suggest an algebra-valued target theory retaining $u$ and the worldline coordinate. Deriving its full action remains open. 

An interesting feature of this vacuum is the following.  In this vacuum the unit vector $\hat\tau$ ranges over a nine-dimensional hyperbolic space, modulo the automorphisms of the lattice. The neutral spectrum is independent of $\hat\tau$, while the energies of the charged states depend on it. As $\hat\tau$ is boosted towards a null vector $\delta$ of the lattice, the states with charges $n\delta$ become light, with energies proportional to $|n|e^{-\eta}$ for rapidity $\eta$. Thus the spectrum develops an evenly spaced light tower, as expected when a spatial dimension decompactifies. We thus expect that in this limit, the worldline becomes the worldsheet of the string vacuum in the first row of Table \ref{tab:vacua}. This is a concrete sense in which the null vector, though not a dynamical field of any one vacuum, is a modulus relating different vacua.

\subsection{The point vacuum and matrix models}
\label{sec:point}

The point vacuum in Table \ref{tab:vacua} suggests a connection to matrix models. Take both null vectors entirely within the base,
\bea
\label{eq:pointnull}
v_1=t^1+x^1~, \hspace{0.4 in} v_2=t^2+x^2~.
\eea
Their span $W$ is a maximal isotropic plane in $\R^{2,2}$, so $W^\perp_{\mathrm{base}}/W=0$. After the identification of the null momentum labels, as in Section \ref{sec:nullghosts}, the shared base is therefore a point, while the remaining left-moving quadratic space is $\II_{9,1}\otimes\R$.

In the neutral left-moving zero-mode sector, the two null constraints and their gauge identifications leave $14-2-2=10$ bosonic components, with a quadratic form of signature $(9,1)$. The two null Clifford constraints reduce the 64-component Weyl spinor in $(11,3)$ to a sixteen-component Majorana-Weyl spinor in $(9,1)$; at zero momentum the Dirac equation imposes no further condition. These are the field representations of the zero-dimensional reduction of ten-dimensional $N=1$ super Yang-Mills theory underlying the Lorentzian IKKT model \cite{Ishibashi:1996xs}. This match motivates a matrix interpretation of the point vacuum, though we do not pursue this here. 

Let us also note that there are some closely related field-theory constructions. In particular, Sezgin constructed a supersymmetric Yang-Mills system in $(11,3)$ using two mutually orthogonal null vectors, with a reduction to ordinary super Yang-Mills theory in $(9,1)$ \cite{Sezgin:1997gr}, while Periwal proposed a matrix model obtained by reducing the constrained Nishino-Sezgin system in $(10,2)$ to a point \cite{Periwal:1996zh}. These examples may give a concrete starting point for investigating the relation between the null reductions considered here and matrix descriptions of string theory.

\subsection{Bosonic left-movers and the Niemeier lattices}
\label{sec:niemeier}

Let us finally mention aspects of Figure \ref{fig:ladder}  for the $N=(2,0)$ string, whose left-movers are those of the bosonic string. With one null current the internal lattice is one of the 24 Niemeier lattices, and these give 24 apparently unrelated theories \cite{Ooguri:1991ie,Kutasov:1996vh}. With two null currents the lattice is $\II_{25,1}$, which is unique. For a primitive null lattice vector $v$, the oscillator reduction and the lattice zero-mode quotient of Section \ref{sec:nullghosts} give the one-current construction with lattice $v^\perp/v$. It is a classical fact that the primitive null vectors of $\II_{25,1}$ fall into 24 orbits, one for each Niemeier lattice \cite{conway1999sphere}; for example $v=(0,1,2,\dots,24\,|\,70)$ gives the Leech lattice. The 24 lattice constructions can therefore be organized as the 24 cusps of one larger lattice construction. The role of these cusps in heterotic compactifications to two dimensions has been discussed in \cite{Harrison:2016pmb}. For the $N=(2,1)$ string the analogous lattice is $\II_{9,1}$, which has a single cusp, corresponding to $E_8$.

\section{Discussion}
\label{sec:discussion}

We have studied an $N=(2,1)$ string which needs no null vector, and found that its interactions are organized by a Lie superalgebra formed by its own states. In this final section we summarize what this suggests about string theory at large, and then discuss several possible interpretations and extensions.

\paragraph{Grand unification.} The picture which emerges is a unification of a rather complete kind, or at least a framework for one. The proposed target theory (\ref{eq:TOE}) is a self-dual gauge theory whose gauge algebra is the Lie superalgebra $\g$ of physical states of the left-movers. Every physical state of the left-movers is then at once a generator of $\g$ and a component of the field $\Phi$, and every cubic coupling is a structure constant. The states of the string are, in other words, the gauge fields of a symmetry which they themselves generate. The neutral generators form the Poisson algebra, extended by the four odd coordinates. Its even part contains the area-preserving diffeomorphisms of the plane $\Lambda$, and the corresponding part of the gauge theory is the self-dual supergravity of Section \ref{sec:neutral}. The generators with null charges form further multiplets of sixteen states, and those with timelike charges are the excited states of the string, with arbitrarily high helicity. Diffeomorphisms, ordinary gauge transformations, and the transformations associated with the higher modes are thus not separate symmetries acting on separate fields---they are parts of one gauge algebra acting on its adjoint representation, whose even and odd elements are related by the supercharges of Section \ref{sec:susy}. In this sense the theory is as ``grand unified'' as possible: there is one field, one algebra, and one cubic vertex, and gravity enters on the same footing as everything else. That such a structure is visible at all is due to the fact that every state is massless on the base. In an ordinary string theory an algebra of this kind, if present, would be complicated by the masses of the excited states.

\paragraph{Points, particles, strings, and branes.} The reduction of Section \ref{sec:mechanism} organizes the vacua of the $N=(2,1)$ string by the position of their null vectors relative to the base, and as summarized in Table \ref{tab:vacua} the resulting target spaces have every dimension from zero to four. With one null vector one finds the string and the membrane originally studied by Kutasov and Martinec \cite{Kutasov:1996fp,Kutasov:1996zm,Kutasov:1996vh}; with two, a worldline and a point; and with none, which is possible only for $\sigma=0$, the four-dimensional theory studied here. In each case there are sixteen supercharges, and the neutral states have the field content of maximally supersymmetric Yang-Mills theory reduced to the dimension in question, that is of the worldvolume theory of a brane, a D-particle for the worldline, and the IKKT matrix model for the point. The null vectors act as moduli which connect these vacua, as we saw in Section \ref{sec:worldline}, where boosting a null vector turns the worldline into a worldsheet. The same worldsheet construction thus describes points, particles, strings, and branes. This extends to lower dimensions the picture of \cite{Kutasov:1996vh}, in which strings and membranes descend from a single object, and it is in line with the relation between $N=(2,1)$ strings and the continuum limit of fully compactified matrix theory proposed in \cite{Martinec:1996ea}. 

What distinguishes the vacua is how much of the algebra of the left-movers they display. An algebra of physical states is present in each of them, for instance the fake monster superalgebra in the worldline vacuum. But in the reduced vacua the null vectors force the momenta of a three-point amplitude to be collinear, so that the right-moving factor $k_1Ik_2$ vanishes and the algebra leaves no trace in the cubic interactions. Among these vacua the $(5,5)$ string is the one which makes the symmetry as manifest as possible: no null vector is chosen and nothing is projected out, so that the whole of $\g$, including its dependence on the holomorphic coordinates of the base, appears directly in nonvanishing cubic amplitudes. This suggests using it as a guide to the others. 

\paragraph{Relation to string field theory.} The proposed theory has the general shape of a string field theory, with its field taking values in a space of states of the string, and its only interaction being cubic and defined by an algebraic operation on that space. It is instructive to compare it with the open string field theory of Witten \cite{Witten:1985cc}. There the field is an arbitrary state of the first-quantized string, ghosts included, the kinetic operator is the BRST charge, and the interaction is an associative product defined by gluing half-strings, whose commutators generate the gauge symmetry. Here the field takes values in the physical states of the left-movers, the kinetic operator is the wave operator of the base, which is the same for every state since all are massless there, and the interaction is the Lie bracket of $\g$ multiplied by the self-dual kinematic factor of the right-movers. That a bracket of open-string type should govern a closed string is a consequence of the heterotic structure. The right-movers contribute a single state, so that the whole spectrum is carried by the left-movers, and a chiral half of a closed string is governed by the same vertex operator algebra as an open string.

For a closed string a cubic action is not what one would expect. Covariant closed string field theory is non-polynomial, with elementary vertices of every order organized by a homotopy Lie algebra \cite{Zwiebach:1992ie}. If the proposal (\ref{eq:TOE}) is correct, the $(5,5)$ string is a closed string with an infinite spectrum, including gravity, whose classical field theory is nevertheless cubic, at least in a frame of light-cone type. The simplification is the one familiar from the closed $N=2$ string, whose single field is described by the cubic Pleba\'nski action \cite{Ooguri:1990ww,Ooguri:1991fp}. For open $N=2$ strings a string field theory using Witten's product was formulated by Berkovits and Siegel \cite{Berkovits:1997pq}, and the present proposal may be regarded as a candidate for its heterotic counterpart, with an infinite tower of states.

\paragraph{A brane interpretation.} Kutasov and Martinec interpreted the target-space theory of the standard $N=(2,1)$ string as the worldvolume theory of an ``M-brane'', a $2+2$ dimensional object in $10+2$ dimensions whose null reductions give strings and membranes \cite{Kutasov:1996vh}. The neutral sector of the $(5,5)$ string admits an interpretation of the same kind, with two differences. Its bosonic action is again the determinant action (\ref{eq:action}), but the six scalars have a metric of signature $(3,3)$, so that it describes a brane with worldvolume $\R^{2,2}$ in a ten-dimensional spacetime of signature $(5,5)$, and no null reduction is required. It is thus not the M-brane, but its analogue for the real form $\sigma=0$. A natural candidate for such a brane exists. Two of the type IIB theories of signature $(5,5)$ obtained by timelike dualities \cite{Hull:1998ym} have 3-brane solutions with worldvolume signature $(2,2)$, and hence transverse signature $(3,3)$, which preserve half of the supersymmetry \cite{Hull:1998fh}. The worldvolume theories of generalized D-branes of this kind are reductions of super Yang-Mills theory in $9+1$ or $5+5$ dimensions \cite{Hull:1999mt}. The reduction of the latter to $2+2$ dimensions has precisely the field content of the neutral sector, namely a vector, six scalars with a metric of signature $(3,3)$, and eight fermions, with sixteen supercharges, and its cubic couplings are the left-moving couplings of Section \ref{sec:threepoint}. The target-space theory combines these couplings with the self-dual kinematics of the right-movers and is of Born-Infeld type, as in \cite{Kutasov:1996vh}. It would be interesting to make this identification precise by comparing (\ref{eq:action}) with the worldvolume action of such a brane. The charged tower, on the other hand, has no counterpart in an ordinary worldvolume theory, which has finitely many fields; for it the comparison with string field theory made above seems the more appropriate one.

\paragraph{Ghost branes and orthosymplectic reduction.} The relation to negative-tension or ghost branes is also suggestive \cite{Okuda:2006fb,Dijkgraaf:2016lym}. Ordinary and ghost branes can give rise to supergroup gauge symmetry, with cancellation between sectors of opposite statistics. The null-current quartet has an analogous cancellation. Negative branes have also been argued to produce regions in which spacetime signature changes, relating string theories in exotic signatures, including $(5,5)$ and $(1,9)$ \cite{Dijkgraaf:2016lym,Hull:1998ym,Hull:1999mt}. The three signatures in Table \ref{tab:realforms} are precisely the ten-dimensional signatures admitting Majorana-Weyl spinors.
 This suggests reading the lattice of the $\sigma=0$ theories as $E_8\oplus E_8(-1)\cong\II_{8,8}$, that is as an ordinary $E_8$ torus together with a ``ghost'' $E_8$ torus of the opposite sign, with five null supercurrents. Indeed gauging five mutually orthogonal null vectors of $\II_{8,8}$ leaves $\II_{3,3}$. 

\paragraph{Two-time physics and the choice of reduction.} Another connection is to two-time physics \cite{Bars:1998ph}, in which a particle in $d$ dimensions is described covariantly in $d+2$ dimensions by gauging an $\mathrm{Sp}(2,\R)$ symmetry acting on the phase-space doublet $(X^M,P^M)$. The odd part of $\osp(r,s|2)$ is such a doublet of vectors, and the square of a general odd element contains the $\mathrm{Sp}(2,\R)$ generators $X\cdot X$, $X\cdot P$, and $P\cdot P$. The element $Q_v$ of (\ref{eq:qvmodel}) corresponds to the degenerate case in which the two vectors are proportional. The similarity between null reduction and the constraints of a conformally covariant particle was already noted in \cite{Kutasov:1996vh}.

This connection makes the choice of reduction an interesting physical question. In the present construction the null vector $v$, and hence the odd differential $Q_v$, is specified externally. A more complete formulation might make this choice dynamical or relate different choices within one theory. The lattice constructions of Section \ref{sec:vacua} already organize several vacua by the choice of null directions. Understanding whether that organization extends to their interactions, and how it is compatible with the right-moving complex structure, could clarify the role of the larger orthosymplectic system.

\section*{Acknowledgements}

The author is supported by JSPS KAKENHI Grant No. 26K17152 and through the Inamori Frontier Program at Kyushu University.
The author thanks Ricardo Monteiro and Natalie Paquette for comments and questions on the first version of this paper. The argument for the vanishing of the tree amplitudes in Section \ref{sec:algebra} was pointed out by Ricardo Monteiro.
The author acknowledges extensive help from AI, namely Claude (Anthropic) and ChatGPT (OpenAI). In particular, aside from the discussion in the main text, a number of additional checks of the proposal of Sections \ref{sec:neutral} and \ref{sec:charged}, including 
\begin{itemize}
\item the vanishing of the sum of residues of the left-moving four-point function of Section \ref{sec:example} for further choices of the charges, in which two of the three channels are of level one, or one channel is of level two;
\item the same statement with an external state of level one;
\item the four-point functions of two fermions and two bosons, and of four fermions, with both bosons and fermions of level one exchanged, including the reproduction of the level-one residue of the four-fermion function by the exchange of the 128 bosons of that level;
\item control computations, in which a coefficient of the bracket (\ref{eq:lbracket}) is changed or an exchanged level is omitted, and the cancellations fail,
\end{itemize}
were performed \emph{exclusively} using AI, and have been reproduced by the author only in the sense of rerunning the provided scripts. Since two independent AI systems agreed on these results, the author expects them to be reliable, and regards them as substantial additional evidence for the proposal. But since they have not been fully reproduced, details of them are not included in the paper. The author takes full responsibility for the things actually written in the paper.

\addtocontents{toc}{\protect\setcounter{tocdepth}{1}}
\appendix

\section{Conventions}
\label{app:conventions}

In this appendix we collect our conventions. Signatures are written as (space, time), so that $\R^{p,q}$ has $p$ spacelike and $q$ timelike directions, and $\II_{a,b}$ denotes the even self-dual lattice of signature $(a,b)$, which exists if and only if $a-b$ is a multiple of eight.

\subsection{The base} 
The base $\R^{2,2}$ has coordinates $x^\mu=(t^1,t^2,x^1,x^2)$ and metric $\mathrm{diag}(-1,-1,+1,+1)$. A vector is written as a bispinor,
\bea
\label{eq:bispinor}
k_{\alpha\dot\alpha}=\begin{pmatrix} k^{t^1}+k^{x^2} & k^{t^2}+k^{x^1} \\ k^{t^2}-k^{x^1} & -k^{t^1}+k^{x^2} \end{pmatrix}~, \hspace{0.4 in} k^2=\det k_{\alpha\dot\alpha}~,
\eea
so that a null vector is $k_{\alpha\dot\alpha}=\lambda_\alpha\tilde\lambda_{\dot\alpha}$ with real spinors $\lambda$ and $\tilde\lambda$, whose components we label by $0$ and $1$. With $\langle ij\rangle=\lambda_{i0}\lambda_{j1}-\lambda_{i1}\lambda_{j0}$, and similarly for $[ij]$, one has $2\,k_i\cdot k_j=\langle ij\rangle[ij]$.

\subsection{The complex structure} We take $I$ to have the nonzero components $I^{t^1t^2}=-I^{t^2t^1}=1$ and $I^{x^1x^2}=-I^{x^2x^1}=-1$, and write $k_1Ik_2=I^{\mu\nu}k_{1\mu}k_{2\nu}$. In terms of (\ref{eq:bispinor}) it acts on the undotted index only, and for two null momenta with $\lambda_i=t_i\lambda$ one finds (\ref{eq:rightfactorbranch}). The adapted complex coordinates are
\bea
\label{eq:complexcoordinates}
z^1=x^1+ix^2~, \hspace{0.4 in} z^2=t^1+it^2~, \hspace{0.4 in} ds^2=dz^1d\bz^{\bar1}-dz^2d\bz^{\bar2}=\eta_{a\bar b}\,dz^ad\bz^{\bar b}~,
\eea
so that $\det\eta=-1$, and we take $\epsilon^{12}=\epsilon^{\bar1\bar2}=1$. The holomorphic components of a momentum are defined by $k_\mu x^\mu=k_az^a+k_{\bar a}\bz^{\bar a}$, that is $k_1={1 \over 2}(k_{x^1}-ik_{x^2})$ and $k_2={1 \over 2}(k_{t^1}-ik_{t^2})$. The plane $\Lambda$ of Section \ref{sec:kinematics} is spanned by the vectors $(1,i,0,0)$ and $(0,0,1,i)$, on which $I$ acts as multiplication by $-i$; in spinor variables it consists of the vectors $\mu_\alpha\tilde\lambda_{\dot\alpha}$ with $\mu=(1,i)$. In the light-cone frame of Section \ref{sec:frame} we take $n={1 \over 2}(0,0,1,-i)$ and $m={1 \over 2}(1,-i,0,0)$, which span $\bar\Lambda$, so that $k_+=n\cdot k=k_1$ and $k_\perp=m\cdot k=k_2$. The invariants (\ref{eq:xdef}) have the following simple form for null momenta,
\bea
\label{eq:xspinor}
X(i,j)=-{1 \over 8}\,\langle\bar\mu i\rangle\langle\bar\mu j\rangle\,[ij]~, \hspace{0.4 in} \bar X(i,j)=-{1 \over 8}\,\langle\mu i\rangle\langle\mu j\rangle\,[ij]~,
\eea
with $\langle\mu i\rangle=\mu_0\lambda_{i1}-\mu_1\lambda_{i0}$. Setting $\lambda_i=t_i\lambda$ and comparing with (\ref{eq:rightfactorbranch}) gives (\ref{eq:xxbar}).

\subsection{The left-movers} Vectors of the ten-dimensional space of the left-movers are written $K=(k;p)$, with metric $\eta_{MN}=\mathrm{diag}(-,-,+,+\,|\,+,+,+,-,-,-)$. In an orthonormal basis $e_I$ of the internal space, the basis (\ref{eq:latticebasis}) of the lattice is $\ell_j=e_j+e_{j+3}$ and $\ell'_j={1 \over 2}(e_j-e_{j+3})$. The free fields are normalized as
\bea
\label{eq:freefields}
i\pd X^M(z)\,i\pd X^N(w)\sim{\eta^{MN} \over (z-w)^2}~, \hspace{0.4 in} \psi^M(z)\,\psi^N(w)\sim{\eta^{MN} \over z-w}~,
\eea
and the correlator of exponentials is $\langle\prod_ie^{iK_i\cdot X}(z_i)\rangle=\prod_{i<j}z_{ij}^{K_i\cdot K_j}$ with $z_{ij}=z_i-z_j$. The modes obey $[\alpha^M_k,\alpha^N_l]=k\,\delta_{k+l}\,\eta^{MN}$ and $\{\psi^M_r,\psi^N_s\}=\delta_{r+s}\,\eta^{MN}$. The vertex operator of a state of charge $p$ also contains a cocycle factor, as a result of which the product of two operators of charges $p$ and $q$ acquires a sign $\epsilon(p,q)$. For $p=m_j\ell_j+m'_j\ell'_j$ and $q=n_j\ell_j+n'_j\ell'_j$ we take $\epsilon(p,q)=(-1)^{m'_jn_j}$, which obeys $\epsilon(p,q)/\epsilon(q,p)=(-1)^{p\cdot q}$ and is equal to one on the sublattice $L$ spanned by the $\ell_j$. Since these signs depend only on the charges, their only effect on a correlator of operators of charges $p_1,\dots,p_n$ is to multiply it by the constant $\prod_{i<j}\epsilon(p_i,p_j)$. The formulas of Appendices \ref{app:correlators} and \ref{app:levelone} are written without these signs, for the operators in the order in which they are listed, and the signs are restored below (\ref{eq:factorizationsigns}). The gamma matrices obey $\{\Gamma_M,\Gamma_N\}=2\eta_{MN}$, and in signature $(5,5)$ the Weyl spinors, with sixteen components, may be taken real. For two spinors of the same chirality we write $\bar u_2\Gamma^Mu_1=u_2^TC\,\Gamma^Mu_1$, with $C$ the charge conjugation matrix.

\section{Zero modes of the null ghosts}
\label{app:zeromodes}

In Section \ref{sec:nullghosts} we noted that the null-current constraint does not by itself identify momenta which differ by a multiple of $v$. Here we ask whether this identification can be obtained from the cohomology of $Q_v$ itself, by including the zero mode of $\theta^-$.

Let $\xi^+$ be the zero mode of $c'=\theta^+$, let $u=v\cdot x$, and choose $\tilde u$ so that $v\cdot\pd/\pd x=\pd/\pd\tilde u$. For the usual $(b',c')$ ghosts, the null-current charge acts on zero-mode wavefunctions as $-\xi^+\pd_{\tilde u}$. At ghost number zero, a function is closed if it is independent of $\tilde u$, and nothing is exact. Such functions may still depend on $u$, with the component of momentum along $v$ remaining unconstrained. Thus the null-current constraint alone removes one coordinate, not two.

One may instead add an odd zero mode $\xi^-$ of $\theta^-$ and choose $Q_0\xi^-=u$. This is an additional choice, since the local relation $Q_v\pd\theta^-=\pd(v\cdot X)$ does not fix the action on the constant mode. The charge acting on functions of $(x,\xi^+,\xi^-)$ is then
\bea
\label{eq:qzero}
Q_0=u\,{\pd \over \pd\xi^-}-\xi^+\,v\!\cdot\!{\pd \over \pd x}~.
\eea
This is the finite-dimensional orthosymplectic differential. For smooth functions near $u=0$, the dependence on $u$ is now exact,
\bea
\label{eq:uexact}
f(u)-f(0)=Q_0\Big[\xi^-\,{f(u)-f(0) \over u}\Big]~,
\eea
and the cohomology consists locally of functions on $v^\perp/v$.

The added zero mode must also be compatible with the remaining constraints. The charge (\ref{eq:qzero}) commutes with the left-moving super-Laplacian $\pd\cdot\pd-2\,\pd_{\xi^+}\pd_{\xi^-}$. However, if $v$ has a component $v_{\mathrm{base}}$ in the shared base, its commutator with the ordinary right-moving Laplacian is $2\,(v_{\mathrm{base}}\cdot\pd)\pd_{\xi^-}$. The corresponding oscillator relation is
\bea
\label{eq:obstruction}
\big[\bar L_n,v\!\cdot\!x\big]\propto v\!\cdot\!\bar\alpha_n~.
\eea
Thus this way of adding $\xi^-$ is obstructed if the right-moving constraints are left unchanged.

For a primitive null vector in a purely left-moving lattice, there is no right-moving obstruction, and the global form of the identification is the lattice zero-mode quotient of Section \ref{sec:nullghosts}. This quotient is not implied by position independence alone. The same construction applies to a compactified base when $v$ is a purely left-moving lattice vector.

In the standard $N=(2,1)$ strings the internal directions are Euclidean, so $v$ necessarily has a component in the base and the obstruction is present. This does not imply that these strings retain an extra coordinate: as recalled in Section \ref{sec:nullghosts}, the identification is implemented there by spectral flow, which removes both zero modes without adjoining $\xi^-$.

\section{Vertex operators and correlators}
\label{app:correlators}

Here we give the correlators used in Section \ref{sec:charged}. All momenta in this appendix are projected momenta in the sense of (\ref{eq:projected}), so that the products $K_i\cdot K_j=p_i\cdot p_j$ are integers.

\subsection{Four bosons} The vertex operators of a massless vector state of momentum $K$ and polarization $e$, with $K^2=e\cdot K=0$, are
\bea
\label{eq:vertexoperators}
V^{(-1)}=e^{-\varphi}\,e\!\cdot\!\psi\,e^{iK\cdot X}~, \hspace{0.4 in} V^{(0)}=\big(e\!\cdot\!i\pd X+K\!\cdot\!\psi\,e\!\cdot\!\psi\big)\,e^{iK\cdot X}~,
\eea
in the $-1$ and $0$ pictures, with $\varphi$ the bosonized superghost. The four-point function of Section \ref{sec:example} is
\bea
\label{eq:fourpoint}
A_L(z)=\big\langle cV_1^{(-1)}(z_1)\ cV_2^{(-1)}(z_2)\ cV_3^{(0)}(z_3)\ V_4^{(0)}(z)\big\rangle=z_{13}\,z_{23}\,\prod_{i<j}z_{ij}^{K_i\cdot K_j}\ W(z_i;e_i,K_i)~,
\eea
where the factor $z_{13}z_{23}$ is the product of the correlator $z_{12}z_{13}z_{23}$ of the reparametrization ghosts and the correlator $1/z_{12}$ of the superghosts, and where $W$ is the sum over Wick contractions of the fields $\psi$ and $i\pd X$ among themselves and of $i\pd X$ with the exponentials, the latter contributing $\sum_jK_j^M/(z-z_j)$. Each contraction is a rational function, and hence so is $A_L$. As $z\to\infty$ the product of exponentials tends to $z^{-K_4\cdot K_4}=1$ and $W$ falls off as $z^{-2}$, by transversality of $e_4$ and momentum conservation, so that there is no residue at infinity. The three-point function of two operators in the $-1$ picture and one in the $0$ picture is independent of the positions and equal to $e_1\!\cdot\!e_2\,e_3\!\cdot\!K_1+e_2\!\cdot\!e_3\,e_1\!\cdot\!K_2+e_3\!\cdot\!e_1\,e_2\!\cdot\!K_3$, the Yang-Mills vertex, up to the ordering sign.

\subsection{Two fermions and two bosons} For two fermions it is convenient to place them at the origin and at infinity, where in the $-{1 \over 2}$ picture they create Ramond ground states $|u_1,K_1\rangle$ and $\langle u_2,K_2|$, with $\slashed{K}_iu_i=0$. Between such states the fermions have the expansion $\psi^M(w)=w^{-1/2}\big(\Gamma^M/\sqrt2+\dots\big)$, where the dots stand for the oscillators, with two-point function $\eta^{MN}w_2/(w_1-w_2)$. With the third operator in the $-1$ picture at $w_3$ and the fourth in the $0$ picture at $w$, all half-integer powers of $w_3$ cancel between the ghosts, the superghosts, and the fermion of the third operator, and one finds
\bea
\label{eq:ffbb}
A_L(w)&=&{w_3^{K_3\cdot K_1}\,w^{K_4\cdot K_1}\,(w_3-w)^{K_3\cdot K_4} \over \sqrt2}\,\Big\{\bar u_2\slashed{e}_3u_1\Big[{e_4\!\cdot\!K_1 \over w}+{e_4\!\cdot\!K_3 \over w-w_3}\Big]+{\bar u_2\slashed{e}_3\slashed{K}_4\slashed{e}_4u_1 \over 2w} \nonumber\\ &&\hspace{1.9 in}+{e_3\!\cdot\!K_4\ \bar u_2\slashed{e}_4u_1-e_3\!\cdot\!e_4\ \bar u_2\slashed{K}_4u_1 \over w_3-w}\Big\}~.
\eea
The three channels correspond to the poles at $w=0$, $w=\infty$, and $w=w_3$. At level zero the residue in a channel in which the fourth operator fuses with the fermion $u_i$ is $\bar u_j\slashed{e}_3[u_4,u_i]/\sqrt2$, with $u_j$ the spinor of the other fermion and with the fusion of a boson and a fermion given by
\bea
\label{eq:bosonfermion}
[u_4,u_i]=\Big(e_4\!\cdot\!K_i+{1 \over 2}\,\slashed{K}_4\slashed{e}_4\Big)u_i~,
\eea
which obeys the Dirac equation at momentum $K_4+K_i$.

\subsection{Four fermions} For four fermions in the $-{1 \over 2}$ picture, the four-point function of the spin fields and of the superghosts gives
\bea
\label{eq:ffff}
A_L(z)&=&{z_{12}z_{13}z_{23} \over 2}\prod_{i<j}z_{ij}^{K_i\cdot K_j-1}\,\Big[\big(\bar u_1\Gamma^Mu_2\big)\big(\bar u_3\Gamma_Mu_4\big)\,z_{14}z_{23} \nonumber\\ &&\hspace{1.75 in}-\big(\bar u_1\Gamma^Mu_4\big)\big(\bar u_3\Gamma_Mu_2\big)\,z_{12}z_{34}\Big]~,
\eea
with $z_4=z$. The spin fields and the superghosts separately have branch cuts, $\prod z_{ij}^{-3/4}$ and $\prod z_{ij}^{-1/4}$, which combine into the integer powers in (\ref{eq:ffff}). At level zero the residue in the channel $(4i|jk)$ is $-{1 \over 2}(\bar u_4\Gamma^Mu_i)(\bar u_j\Gamma_Mu_k)$, and the vanishing of the sum of the three residues is the Fierz identity $\Gamma^M_{(\alpha\beta}\Gamma_{M\,\gamma)\delta}=0$.

\section{The first massive level}
\label{app:levelone}

In this appendix we describe the bosonic states of level one, and their couplings to pairs of massless states.

\subsection{States} A Neveu-Schwarz state of level one has momentum $Q$ with $Q^2=-2$ and is of the form
\bea
\label{eq:leveloneansatz}
|a,b,c\,;Q\rangle=\Big[a_M\,\psi^M_{-3/2}+b_{MN}\,\alpha^M_{-1}\psi^N_{-1/2}+{1 \over 6}\,c_{MNR}\,\psi^M_{-1/2}\psi^N_{-1/2}\psi^R_{-1/2}\Big]|Q\rangle~,
\eea
with $c$ antisymmetric, so that there are $10+100+120=230$ such states. The physical-state conditions, from $G_{1/2}$ and $G_{3/2}$, are
\bea
\label{eq:levelonephysical}
a_M+b_{MN}Q^N=0~, \hspace{0.4 in} b_{MN}-b_{NM}+Q^Rc_{RMN}=0~, \hspace{0.4 in} a\!\cdot\!Q+b^M{}_M=0~.
\eea
These leave 174 states, of which 46 are null, being of the form $G_{-1/2}$ or $G_{-3/2}$ acting on states of lower level. The remaining 128 are the bosons of the multiplet of Table \ref{tab:multiplets}. In the norm $a\cdot a+b_{MN}b^{MN}+{1 \over 6}c_{MNR}c^{MNR}$ they have signature $(64,64)$.

\subsection{Fusion of two massless bosons} For two massless vectors with $K_1\cdot K_2=-1$, the bracket (\ref{eq:gbracket}) is the state (\ref{eq:leveloneansatz}) of momentum $Q=K_1+K_2$ with
\bea
\label{eq:levelonebracket}
a&=&(e_1\!\cdot\!e_2)K_1-(K_1\!\cdot\!e_2)e_1~, \hspace{0.4 in} c=K_1\wedge e_1\wedge e_2~, \nonumber\\ b&=&e_1\otimes e_2+K_1\otimes\big[(e_1\!\cdot\!K_2)e_2+a\big]~,
\eea
where $(K_1\wedge e_1\wedge e_2)_{MNR}$ is the sum of the six signed products, without a factor of $1/3!$. One checks that it obeys (\ref{eq:levelonephysical}). Note that (\ref{eq:levelonebracket}) is the contour product of the operators without cocycle factors. Under the exchange of the two vectors it is symmetric, up to a null state, and the bracket of $\g$ is obtained from it by multiplying by the cocycle sign $\epsilon(p_1,p_2)$ of Appendix \ref{app:conventions}, which makes it antisymmetric. As the eight polarizations of each of the two vectors are varied, these states span a space of dimension 64, on which the norm has signature $(32,32)$. The corresponding vertex operator in the $-1$ picture is $e^{-\varphi}\big(a\!\cdot\!\pd\psi+b_{MN}\,i\pd X^M\psi^N+{1 \over 6}c_{MNR}\,\psi^M\psi^N\psi^R\big)e^{iQ\cdot X}$, and it is this operator which we used for the external leg of level one in Section \ref{sec:algebra}.

\subsection{The invariant form} The inner product appearing in (\ref{eq:factorization}) pairs states of momenta $Q$ and $-Q$. At level zero it is $e\cdot e'$, and the bracket of two vectors with $K_1\cdot K_2=0$ has the polarization $(e_1\!\cdot\!K_2)\,e_2-(K_1\!\cdot\!e_2)\,e_1+(e_1\!\cdot\!e_2)\,K_1$. At level one it is
\bea
\label{eq:invariantform}
\big(|a,b,c\,;Q\rangle,|a',b',c'\,;-Q\rangle\big)=-a\!\cdot\!a'+b_{MN}b'^{MN}-{1 \over 6}\,c_{MNR}c'^{MNR}~.
\eea
This bilinear pairing relates opposite momenta and differs from the same-momentum oscillator norm above. The displayed relative signs are fixed here by matching the sphere factorization conventions; they reproduce the independent two-boson and two-fermion exchange calculations. With these definitions, the precise form of (\ref{eq:factorization}) is
\bea
\label{eq:factorizationsigns}
\Res_{z\to z_i}A_L(z)=\sigma_i\,\big([u_4,u_i],[u_j,u_k]\big)~, \hspace{0.4 in} \sigma_i=(-1)^{\sum_{j<i}K_4\cdot K_j}~,
\eea
with $(i,j,k)$ a cyclic permutation of $(1,2,3)$. The signs $\sigma_i$ arise from moving the fourth operator past the others in the absence of cocycle factors. When the cocycle factors of Appendix \ref{app:conventions} are included, the correlator is multiplied by $\prod_{i<j}\epsilon(p_i,p_j)$, the brackets $[u_4,u_i]$ and $[u_j,u_k]$ by $\epsilon(p_4,p_i)$ and $\epsilon(p_j,p_k)$, and the inner product of two states of charges $P$ and $-P$ by $\epsilon(P,-P)=(-1)^{P^2/2}$. The product of these four signs is $\sigma_i$, so that with the cocycle factors included (\ref{eq:factorization}) holds with a plus sign in all three channels.

\subsection{Couplings to two fermions} Two massless fermions with spinors $u_1$ and $u_2$ and momenta $K_1$ and $K_2$ couple to the state (\ref{eq:leveloneansatz}) of momentum $-(K_1+K_2)$ through
\bea
\label{eq:fermionpair}
C(u_2,u_1;a,b,c)={1 \over \sqrt2}\,\Big[-{1 \over 2}\,\bar u_2\slashed{a}u_1+K_{1M}\,b^{MN}\,\bar u_2\Gamma_Nu_1+{1 \over 12}\,c^{MNR}\,\bar u_2\Gamma_M\Gamma_N\Gamma_Ru_1\Big]~.
\eea
The residue of (\ref{eq:ffbb}) at a pole at which a level-one boson is exchanged is minus this coupling, evaluated on the state (\ref{eq:levelonebracket}) built from the two bosons. Regarded as linear functions on the physical states of level one, the couplings (\ref{eq:fermionpair}) for the $8\times8$ pairs of physical spinors span a space of dimension 64, those of the pairs of bosons, obtained from (\ref{eq:invariantform}) and (\ref{eq:levelonebracket}), another space of dimension 64, and the two together a space of dimension 128. All of them vanish on the null states. Finally, the residue of the four-fermion function (\ref{eq:ffff}) at a level-one pole is equal to minus the sum over the 128 bosons of the product of two couplings (\ref{eq:fermionpair}), with the inverse of the form (\ref{eq:invariantform}) as propagator. We have not constructed the 128 fermions of level one explicitly; their exchange is included in the correlator (\ref{eq:ffbb}), and they contribute to the sums of residues described in Section \ref{sec:algebra}.

\bibliographystyle{ytamsalpha}
\bibliography{refs}

\end{document}